\documentclass{aa}  

\usepackage{natbib}
\usepackage{graphicx}
\usepackage{txfonts}

\usepackage[dvipsnames]{xcolor}
\usepackage{tikz}
\usetikzlibrary{bayesnet}

\usepackage{diagbox}
\usepackage{subcaption}

\newcommand{\de}{\text{d}}
\newcommand{\Msun}{\text{M}_{\odot}}
\newcommand{\kmsec}{\text{km}\,\text{s}^{-1}}

\newcommand{\kpc}{\text{kpc}}
\newcommand{\pc}{\text{pc}} 
\newcommand{\yr}{\text{yr}}

\newcommand{\Msunppcc}{\Msun\pc^{-3}}

\newcommand{\mas}{\text{mas}}

\newcommand{\popp}{\boldsymbol{\Psi}}

\newcommand{\data}{\boldsymbol{d}}

\newcommand{\sigm}{\text{sigm}}

\defcitealias{PaperI}{Paper~I}
\defcitealias{PaperII}{Paper~II}
\defcitealias{PaperIV}{Paper~IV}

\usepackage{hyperref}
\begin{document}

   \title{Weighing the Galactic disk using phase-space spirals \\ III. Probing distant regions of the disk using the \emph{Gaia} EDR3 proper motion sample}
   \titlerunning{Weighing the Galactic disk using phase-space spirals III}

   \author{A. Widmark
          \inst{1}
          \and
          C.F.P. Laporte
          \inst{2}
          \and
          G. Monari
          \inst{3}
          }

   \institute{Dark Cosmology Centre, Niels Bohr Institute, University of Copenhagen, Jagtvej 128, 2200 Copenhagen N, Denmark\\
   \email{axel.widmark@nbi.ku.dk}
   \and
   Institut de Ci\`encies del Cosmos (ICCUB), Universitat de Barcelona (IEEC-UB), Mart\'i i Franqu\`es 1, 08028 Barcelona, Spain
   \and
   Universit\'e de Strasbourg, CNRS UMR 7550, Observatoire astronomique de Strasbourg, 11 rue de l'Universit\'e, 67000 Strasbourg, France
    }

   \date{Received Month XX, XXXX; accepted Month XX, XXXX}

 
  \abstract{
  We have applied our method to weigh the Galactic disk using phase-space spirals to the proper motion sample of \emph{Gaia}'s early third release (EDR3). For stars in distant regions of the Galactic disk, the latitudinal proper motion has a close projection with vertical velocity, such that the phase-space spiral in the plane of vertical position and vertical velocity can be observed without requiring that all stars have available radial velocity information. We divided the Galactic plane into 360 separate data samples, each corresponding to an area cell in the Galactic plane in the distance range of 1.4--3.4 kpc, with an approximate cell length of 200--400 pc. Roughly half of our data samples were disqualified altogether due to severe selection effects, especially in the direction of the Galactic centre. In the remainder, we were able to infer the vertical gravitational potential by fitting an analytic model of the phase-space spiral to the data. This work is the first of its kind, in the sense that we are weighing distant regions of the Galactic disk with a high spatial resolution, without relying on the strong assumptions of axisymmetry. Post-inference, we fitted a thin disk scale length of $2.2\pm 0.1~\kpc$, although this value is sensitive to the considered spatial region. We see surface density variations as a function of azimuth of the order of 10--20~\%, which is roughly the size of our estimated sum of potential systematic biases. With this work, we have demonstrated that our method can be used to weigh distant regions of the Galactic disk despite strong selection effects. We expect to reach even greater distances and improve our accuracy with future \emph{Gaia} data releases and further improvements to our method.
  }


   \maketitle
%

\section{Introduction}\label{sec:intro}

The dynamics of stars can be related to the gravitational potential that they inhabit via the collisionless Boltzmann equation. For systems in a steady state with certain symmetry properties (typically spherical or axisymmetric) it is possible to find solutions to the phase-space distribution of a stellar tracer population, either through distribution function modelling or via the moments of the Boltzmann equation \citep{Kapteyn1922,Oort1932,1984ApJ...287..926B,1984ApJ...276..169B,KuijkenGilmore1989a,Creze1998,HolmbergFlynn2000,Bienayme:2005py,BT2008,doi:10.1111/j.1365-2966.2012.21608.x,bovyrix13,2020A&A...643A..75S,2020MNRAS.495.4828G,2021A&A...646A..67W}. Given the relatively quiet conditions necessary to form disk galaxies, the assumption of equilibrium for near-equilibrium systems has been widely and favourably applied to the Milky Way and other galaxies \citep{mcmillan11, binney11}. Our place in the Milky Way makes it ideal to accurately measure its gravitational potential and mass distribution, since it is the only system where we have access to full six-dimensional phase space information, from its inner regions all the way to its outermost edge \citep[e.g.][]{deason21}. A precise and robust measurement of the gravitational potential is crucial for our general understanding of the Milky Way \citep{1998MNRAS.294..429D,Klypin:2001xu,Widrow:2008yg,2010A&A...509A..25W,mcmillan11,2014ApJ...794...59K,2017MNRAS.465..798C,2017MNRAS.465...76M}, and also for probing its distribution of dark matter \citep{Read2014,2020MNRAS.494.6001N,2020MNRAS.494.4291C,2020ApJ...894...10L,2021RPPh...84j4901D}. The local dark matter density is of fundamental importance for direct and indirect dark matter detection experiments \citep{Jungman:1995df,2015PrPNP..85....1K}. In a broader sense, dark matter's gravitational signatures, studied via stellar dynamics and gravitational lensing, is one of the most competitive avenues for constraining its thus far elusive particle nature \citep{bertone18, ferreira20}. The \emph{Gaia} satellite has been instrumental to this field, pushing the size of the astrometric sample from a few hundred thousand stars \citep{hipparcos} to roughly two billion \citep{Gaia18DR2_summary}.

With \emph{Gaia}, it has become evident that the Milky Way disk is not in a steady state; rather, it is a dynamical system with clear time-varying features, for example in the form of radial and vertical asymmetries \citep{widrow12, williams13,gaia_kinematics}. This is expected from a theoretical perspective, considering that Milky Way-like galaxies do undergo interactions with satellite galaxies which can warp and heat the Galactic disk, induce spiral arms and corrugations \citep{velazquez99, villalobos08, purcell11, gomez13,2020MNRAS.499.5623Q}, which are also in interplay with giant molecular clouds \citep{d'onghia13}, as well as induce bar formation \citep{hohl71}. In fact, such types of dynamical features were seen tentatively also in the pre-\emph{Gaia} era (e.g. \citealt{minchev09}; now confirmed with \citealt{ramos18}). The broken steady state is perhaps most clearly exemplified by the recently discovered phase-space spiral \citep{2018Natur.561..360A}, seen in the plane of vertical position and vertical velocity for stars in the solar neighbourhood. It is probably the remnant of a perturbation event a few hundred million years ago, although the precise time is highly uncertain (see e.g. \citealt{2019MNRAS.484.1050D}). Pre-\emph{Gaia} self-consistent models of the Milky Way's interaction with a Sagittarius-like satellite \citep{laporte18} have been used to interpret the formation of the phase-space spiral as a global phenomenon \citep{laporte19}, demonstrating that the spiral is present many kilo-parsec beyond the solar neighbourhood and that it has a similar shape for stars of essentially all ages (at least in the range of roughly 1--9 billion years). Its global nature was further demonstrated in subsequent works \citep[e.g.][]{li20, antoja21}. This indicates that the perturbation should probably have a recent origin, ruling out bar buckling \citep{khoperskov19} which would also violate constraints on the Galactic bar's chemical structure \citep{ness13, debattista19}.

In order to deepen our understanding of the Galaxy, it seems all the more fruitful and necessary to go beyond the common assumptions of symmetry and a steady state (what is referred to as an Ideal Galaxy in \citealt{2021RPPh...84j4901D}). There has been significant effort in testing the traditional steady state based methods against simulations, in order to control for the systematic biases that might arise due to the breaking of symmetry or a steady state. Generally, such methods perform well (\citealt{2019ApJ...879L..15H}; \citealt{2020A&A...643A..75S}; \citealt{2022MNRAS.511.1977S}), but in principle we can go further than that. As an example, \cite{2021MNRAS.503.1586L} apply steady state modelling to the solar neighbourhood but use the phase-space spiral as a consistency check, by comparing a spiral model with the residual that emerges in their inferred equilibrium distribution. There is also the possibility to extract information directly from time-varying dynamical structures, which this work is an example of.

This article is the third part of a longer series about a new method for weighing the Galactic disk, in which the vertical gravitational potential is inferred from the shape of the phase-space spiral in the plane of vertical position and vertical velocity. A way to think about how this method gains its power of inference is to consider the following. Given that the spiral has winded into its current shape in a fairly stable disk, its spiral angle, defined in Eq.~\eqref{eq:angle_of_time}, is a smooth and monotonic function of vertical energy. In other words, if we trace a curve along the arm of the spiral, from the inside out in the vertical phase-space plane, the vertical energy should be smoothly and strictly increasing. Given this property, the shape of this curve places a very strong constraint on the vertical gravitational potential. This property is expected to hold even though the stars that collectively make up the spiral perturbation have varying dynamical histories and Galactocentric guiding radii. Given that a well-defined single-armed spiral is visible, it is difficult and contrived to imagine a scenario where the property of a strictly increasing vertical energy is broken.

In the two previous papers in this series---\cite{PaperI} and \cite{PaperII}, henceforth referred to as \citetalias{PaperI} and \citetalias{PaperII}---we have tested our method on one-dimensional simulations and applied our method to the immediate solar neighbourhood using the radial velocity sample of \emph{Gaia}'s early instalment of its third data release (EDR3). In a subsequent fourth article, we have also tested our method on a high-resolution three-dimensional $N$-body simulation---\cite{PaperIV}, henceforth referred to as \citetalias{PaperIV}---which further motivates and supports this work and its results. Our method produced accurate results for the rich and complex dynamics of an externally perturbed three-dimensional disk galaxy. We also demonstrated our method's robustness with respect to severe and unknown spatially dependent selection effects, as well as a biased height of the disk mid-plane.

In this third article, we applied our method to distant regions of the Galactic disk, using the \emph{Gaia} EDR3 proper motion sample. While radial velocity information is essential for seeing the phase-space spiral in the immediate solar neighbourhood, it is less important when observing the spiral at a distance of a few kilo-parsec. The reason is that disk stars at such distances have a small Galactic latitude ($b\simeq 0~\deg$), such that the proper motion in the latitudinal direction has a close projection to vertical velocity (proportional to $\cos b$), while the contribution coming from the radial velocity component is small (proportional to $\sin b$). We constructed 360 separate data samples, by dividing the Galactic disk into different area cells in the directions parallel to the Galactic plane, with a distance bin length of 200 pc in the range 1.4--3.4~kpc and a Galactic longitude bin length of ten degrees. This analysis is the first of its kind, in the sense that we are weighing distant regions of the Galactic disk with a high spatial resolution in the directions parallel to the Galactic plane. Because all data samples are fitted individually, our inference is not subject to otherwise commonly made assumption about the large-scale spatial structure of the Galactic potential (mainly the assumptions of axisymmetry and a disk matter density that decays exponentially with Galactocentric radius). While we constructed 360 data samples, we could only apply our method to roughly half of them, mainly due to severe selection effects in the general direction of the Galactic centre. In order to extract the shape of the spiral in the remaining spatial volume, we modified the method used in this work as compared to \citetalias{PaperI} and \citetalias{PaperII} by adding a simple extinction model as a function of spatial position. Unlike traditional methods that are based on the assumption of a steady state, our modelling of selection effects is not required to be very precise, but only good enough in order to robustly extract the shape of the phase-space spiral.

This article is structured as follows. In Sect.~\ref{sec:definitions}, we define a coordinate system and a few other key quantities. We describe the data sample construction in Sect.~\ref{sec:data} and our model of inference in Sect.~\ref{sec:model}. In Sect.~\ref{sec:results}, we present our results. In Sects.~\ref{sec:discussion} and \ref{sec:conclusion}, we discuss and conclude.

\section{Coordinate system and gravitational potential}\label{sec:definitions}

In this article, we use a Cartesian system of coordinates centred on the Sun's location and rest frame, whose spatial coordinates $\boldsymbol{X} \equiv \{X,Y,Z\}$ correspond to the direction of the Galactic centre, the direction of Galactic rotation and the direction of Galactic north, respectively. The time derivatives in the three directions give the velocities $\boldsymbol{V} \equiv \{U,V,W\}$. How these phase-space coordinates are related to the \emph{Gaia} observables is described in Appendix~\ref{app:coord_trans}.

The height, also referred to as vertical position, with respect to the Galactic plane is written as
\begin{equation}\label{eq:Z_to_z}
    z = Z + Z_\odot,
\end{equation}
where $Z_\odot$ is the height of the Sun relative to the Galactic mid-plane. The velocity in vertical direction in the rest frame of the Galactic disk is written as
\begin{equation}\label{eq:W_to_w}
    w = W + W_\odot,
\end{equation}
where $W_\odot$ corresponds to the Sun's velocity relative to the disk's bulk motion. In the immediate viscinity of the Sun, we have that $W_{\odot,\text{local}} = 7.25~\kmsec$ (from e.g. \citealt{2010MNRAS.403.1829S}). However, because we are studying distant parts of the disk over which the mode of the vertical velocity distribution can vary, we let $W_\odot$ be a free parameter, fitted individually for the respective regions of Galactic disk where we applied our method. Ideally, $Z_\odot$ should have been made a free parameter as well, but this quantity cannot be robustly inferred for distant disk regions due to severe selection effects. Instead, we assume $Z_\odot$ to be given directly by the height of the Sun in the immediate solar neighbourhood; in other words, we approximated the Galactic disk mid-plane to be perfectly flat within the considered distances ($<3.4~\kpc$). The Sun's height is typically evaluated to lie in the range 0--20 pc. A broken Galactic plane symmetry is indicated by the fact that lower values for $Z_\odot$ are preferred in more local studies (e.g. \citealt{Buch:2018qdr,2021A&A...646A..67W,2021A&A...649A...6G}), as opposed to studies that reach several kilo-parsec in distance (e.g. \citealt{Juric:2005zr,2017MNRAS.468.3289Y,BovyAssym}). For this work, we chose a fixed value of $Z_\odot = 10~\pc$.

In our model, the vertical gravitational potential is written as
\begin{equation}\label{eq:phi}
\begin{split}
    \Phi(z\,|\,\rho_h) = & \sum_{h=1}^{4}
    4 \pi G \rho_h (2^{h-1} \times 200~\pc)^2 \\
    & \times \log\Bigg[\cosh\Bigg(\dfrac{z}{2^{h-1} \times 200~\pc}\Bigg)\Bigg],
\end{split}
\end{equation}
which corresponds to a total matter density consisting of a sum of four matter density components with different scale heights of $\{200,400,800,1600\}~\pc$ (this is found via the Poisson equation, see \citetalias{PaperII} for further details). Using this functional form, the gravitational potential is free to vary in shape, and is flexible enough to emulate models for the total matter density of the solar neighbourhood (e.g. \citealt{2015ApJ...814...13M} and \citealt{Schutz:2017tfp}).

The vertical energy per mass is given by the vertical gravitation potential and vertical kinetic energy per mass, according to
\begin{equation}
    E_z = \Phi(z \, | \, \rho_h) + \frac{w^2}{2}.
\end{equation}
However, throughout this paper, we refer to any ``energy per mass'' as simply ``energy'', for shorthand.

Although it does not directly enter our model of inference, we used the Galactocentric cylindrical radius when interpreting our results. This quantity is given by
\begin{equation}
    R = \sqrt{(R_\odot - X)^2+Y^2},
\end{equation}
where we assumed a solar position of $R_\odot = 8178~\pc$ \citep{2019A&A...625L..10G}.

\section{Data}\label{sec:data}

The main difference between this work and previous articles in this series is that we have used the parallax and proper motion information of \emph{Gaia} EDR3 without requiring the availability of radial velocities. When extracting the shape of the spiral in the vertical phase-space plane, we need the vertical velocity ($W$), while other velocity components ($U$ and $V$) are largely irrelevant. For distant regions of the Galactic disk, the Galactic latitude is close to zero, such that the proper motion in the latitudinal direction has a close projection to vertical velocity. While this approximation would not be feasible in the immediate solar neighbourhood, it is reasonable for disk stars at kilo-parsec distances. This is discussed more carefully in Sects.~\ref{sec:RV_assignment} and \ref{sec:biases} below.

The radial velocity information of \emph{Gaia} EDR3 was included when available, and was also supplemented with legacy spectroscopic surveys (compiled in \citealt{SD18}, including LAMOST DR3 \citealt{Deng12}, GALAH DR2 \citealt{buder18}, RAVE DR5 \citealt{kunder17}, APOGEE DR14 \citealt{abolfathi18}, SEGUE \citealt{Yanny09}, and GES DR3 \citealt{Gilmore12}). Similar to in \citetalias{PaperII}, a supplementary radial velocity was used if the \emph{Gaia} radial velocity information was missing or had an uncertainty larger than $3~\kmsec$. When there were radial velocity measurements from several supplementary surveys, we used the measurement with the smallest uncertainty. Additionally, we disregarded the supplementary radial velocity information altogether for stars with discrepant measurement values in the supplementary surveys; this was the case if two separate measurement, each with a precision of at least $\sigma_{\text{RV}} < 5~\kmsec$, had a statistical tension of greater than $2.5\sigma$ between them.

The data used to construct the data samples of this work consisted of parallax and its associated uncertainty ($\varpi$ and $\sigma_\varpi$), Galactic longitude and latitude ($l$ and $b$), proper motions ($\mu_b$, $\mu_l$), the astrometric renormalised unit weight error (RUWE), radial velocity and its associated uncertainty ($v_{\text{RV}}$ and $\sigma_{\text{RV}}$, only for a subset of stars). We corrected for the parallax zero-point offset in \emph{Gaia} EDR3 according to the bias function described in \cite{edr3_parallax_offset}. In order to compute this, we also included the data columns
\texttt{phot\_g\_mean\_mag},
\texttt{nu\_eff\_used\_in\_astrometry},
\texttt{pseudocolour},
\texttt{ecl\_lat},
\texttt{astrometric\_params\_solved}.

\subsection{Data cuts}\label{sec:data_cuts}

When constructing our data samples, we made cuts in data quality by requiring $\text{RUWE}<1.4$ and $\sigma_{\varpi}<0.05~\mas$. We included the radial velocity information only for stars with $\sigma_{\text{RV}}<3~\kmsec$. These cuts in data quality set an effective limit in apparent magnitude, due to dimmer sources having worse astrometric precision; the number of included sources drop quickly for \emph{Gaia} $G$-band apparent magnitudes in range 16--17. These quality cuts also induce strong spatially dependent selection effects, because of how dust extinction and stellar crowding affect \emph{Gaia}'s completeness and astrometric precision.

We divided the disk plane into area cells which are 10 degrees wide in Galactic longitude ($l$) and 200 pc in terms of distance parallel to the Galactic plane ($\sqrt{X^2+Y^2}$), ranging from 1400 to 3400 pc. This amounts to a total number of $36 \times 10 = 360$ separate area cells.

We also performed a cut in the longitudinal proper motion divided by parallax ($\mu_l / \varpi$), which was performed separately for each individual data sample. We excluded any star for which $\mu_l / \varpi$ was more than one standard deviation away from the mode of that data sample's distribution. This was done in order to exclude stars that are evidently not moving in unison with the bulk of the stellar disk, such as stars with highly eccentric orbits. This phase-space cut is not ideal, in the sense we are cutting in only one out of two velocities parallel to the Galactic plane; regardless, making this crude cut is helpful in isolating the shape of the phase-space spiral.

\begin{figure}
	\includegraphics[width=1.\columnwidth]{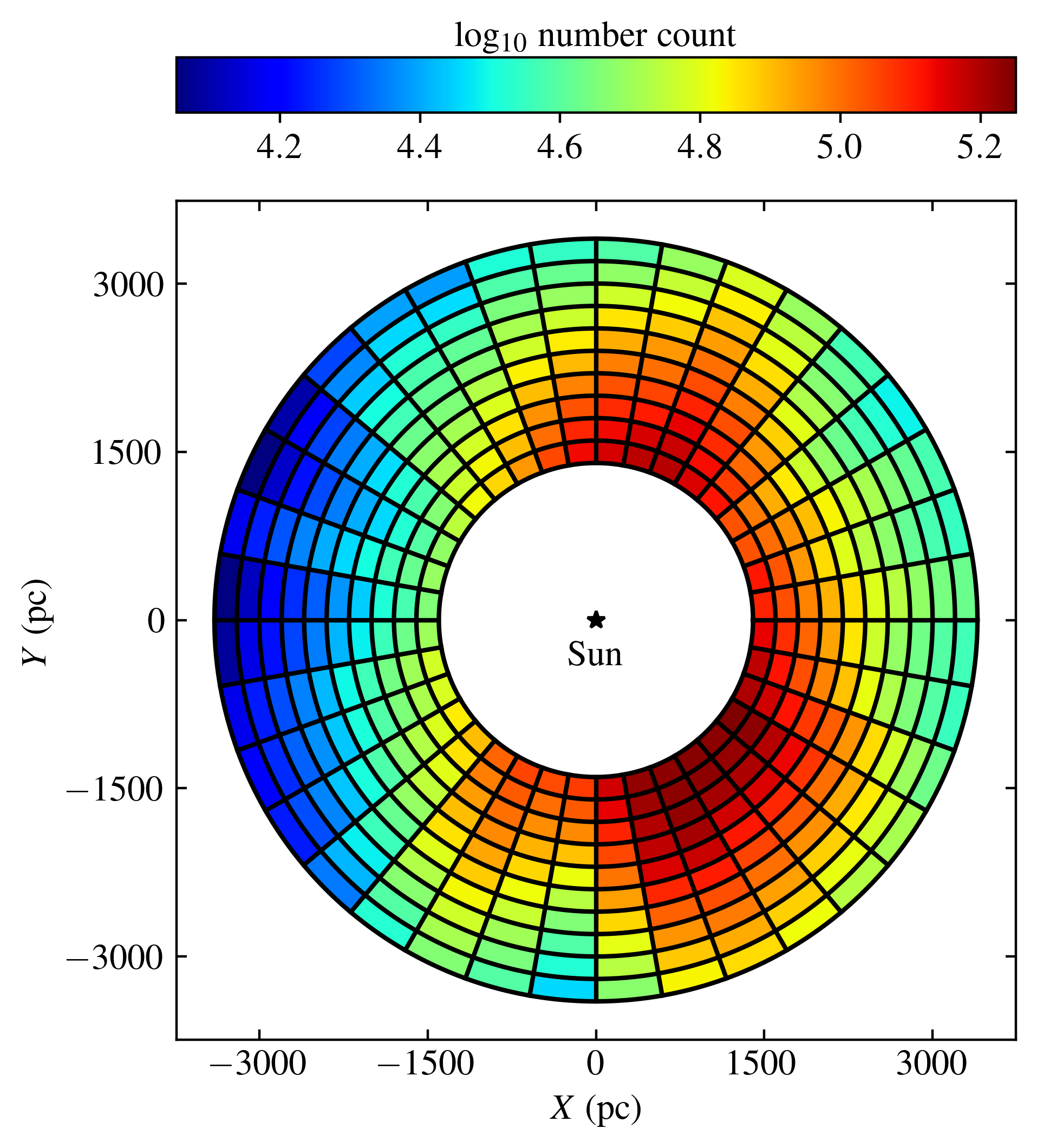}
    \caption{Stellar number counts in our 360 data samples, shown in their respective locations in the $(X,Y)$-plane. The Galactic centre is located to the right, while the direction of Galactic rotation is upwards.}
    \label{fig:number_count_area_cells}
\end{figure}

The area cells in the $(X,Y)$-plane, and their respective stellar number counts, are visible in Fig.~\ref{fig:number_count_area_cells}. However, due to severe selection effects, we could only analyse about half of those data samples using our phase-space spiral method. This is discussed further in Sect.~\ref{sec:results}.

\subsection{Assigning radial velocities}
\label{sec:RV_assignment}

A star's vertical velocity in the solar rest frame is equal to
\begin{equation}\label{eq:vertical_vel}
	W = k_\mu \frac{\mu_b}{\varpi} \cos{b} + v_{\text{RV}} \sin{b}.
\end{equation}
It is the sum of the projected velocities in the latitudinal and line-of-sight directions, which are proportional to $\sin b$ and $\cos b$, respectively. For distant regions of the Galactic disk, the latitude ($b$) is close to zero, such that the proper motion term dominates in the above equation. For this reason, we can approximate the vertical velocity to decent precision even if we only have proper motion information.

\begin{figure}
	\includegraphics[width=1.\columnwidth]{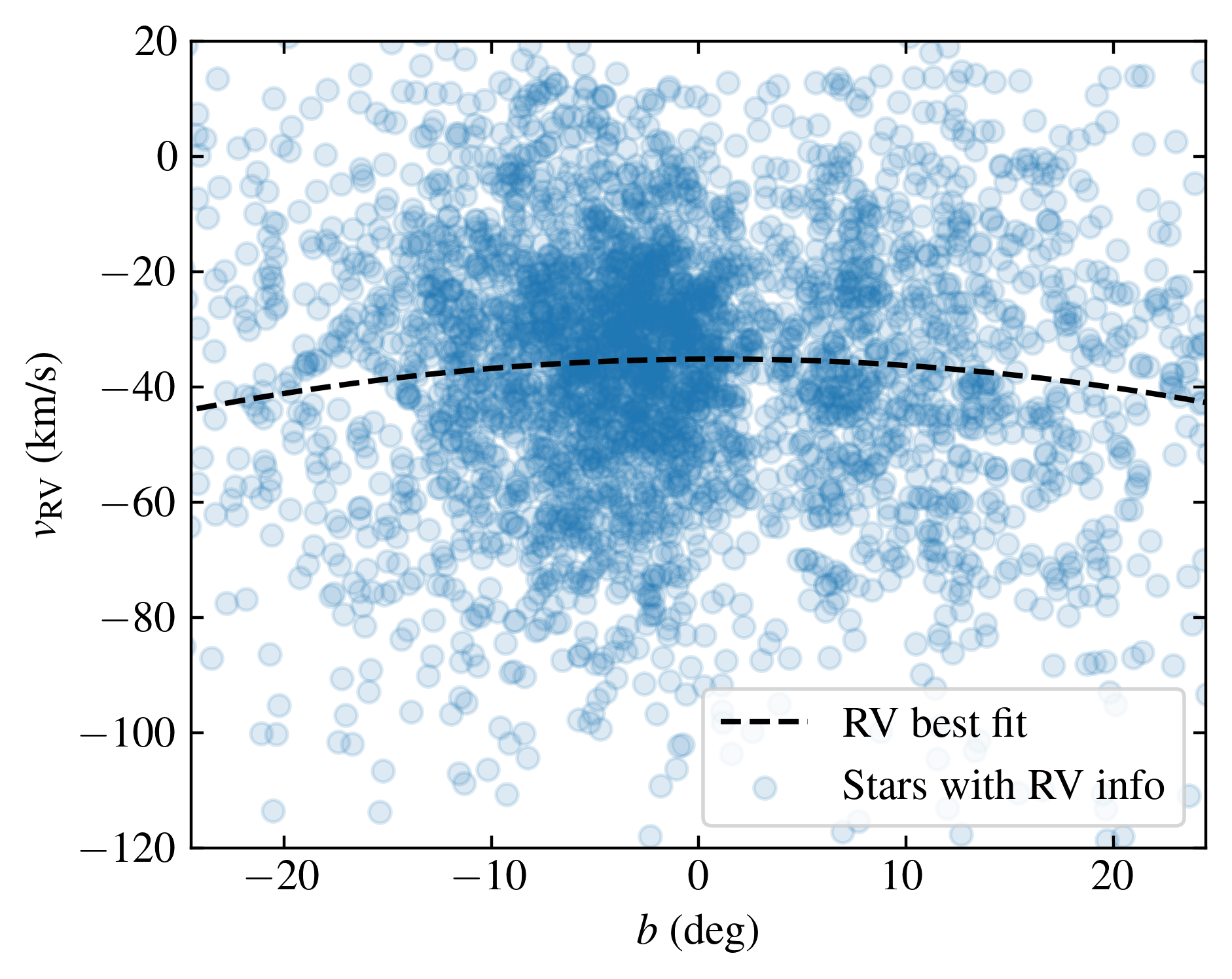}
    \caption{Radial velocity ($v_\text{RV}$) as a function of Galactic latitude ($b$) for the data sample with $l\in[90,100]~\deg$ and $\sqrt{X^2+Y^2}\in [2200, 2400]~\pc$. The dashed line corresponds to the fitted function and the scatter point correspond to stars with available radial velocity information. The range in $b$ on the horizontal axis corresponds to a range in height spanning roughly $Z\in[-1,1]~\kpc$ at the relevant distance.}
    \label{fig:vRV_of_b}
\end{figure}

For the stars in our data samples that were missing radial velocity information, we assigned these missing values. We did so by estimating the group velocity in the line-of-sight direction, using the sub-sample of stars with available radial velocities. For each respective data sample, we fitted a second degree polynomial of $v_{\text{RV}}$ as a function of $b$, and assigned the missing radial velocity values according to that function. The second degree polynomial was completely free to vary in this fit and not required to be symmetric with respect to for example $b=0~\deg$; as such, this fit is independent of the assumed value for $Z_\odot$. Because the data samples were fairly small in the $X$ and $Y$ directions, fitting this simple function was enough to capture the most important dependency of $v_{\text{RV}}$. This interpolation is shown in Fig.~\ref{fig:vRV_of_b} for a representative data sample, which contained a subset of 3528 stars with radial velocity information. A discussion about possible biases that can arise from this procedure is found below, in the second to last paragraph of Sect.~\ref{sec:biases}.

\subsection{Observational and systematic uncertainties}
\label{sec:biases}

A useful way to understand how our method can infer the vertical gravitational potential is that the shape of the phase-space spiral in the $(z,w)$-plane informs us about what stars have similar vertical energies (see e.g. figure 1 in \citetalias{PaperI}). Hence, the gravitational potential difference between stars can be known, to the extent that the difference in vertical kinetic energy ($w^2/2$) is known as well. If the height or vertical velocity information is biased, this will propagate into a bias in terms of the inferred gravitational potential. The most significant sources of potential bias are discussed below.

In this work, we made the approximation that the Galactic plane is perfectly flat in the studied volume, such that Eq.~\eqref{eq:Z_to_z} holds for all data samples. However, the height of the Galactic mid-plane most likely varies with $X$ and $Y$. In order to ameliorate this systematic uncertainty we would need a deeper and more detailed understanding of the Galactic disk's spatial structure, which is unfortunately beyond the scope of this work. Due to incompleteness effects and other data systematics, inferring this spatial dependence is difficult, especially given the data quality cuts that were used in this work. As we saw in our analysis of the immediate solar neighbourhood in \citetalias{PaperII}, varying $Z_\odot$ between the values $\{0,10,20\}~\pc$ affected the inferred value of $\Phi(400~\pc)$ by only a few per cent. In this work, the disk mid-plane is probably confined to vary by a few tens of parsecs within the studied volume; looking at the larger scale warp of the Galactic disk, as studied using Cepheids in for example \cite{2019AcA....69..305S} and \cite{2019NatAs...3..320C}, vertical variations of the Galactic mid-plane seem to reach scales of $100~\pc$ and above only at distances around 5 kpc from the Sun. To the extent that the disk mid-plane does vary, it will bias our result by distorting the shape of the inferred phase-space spiral. Roughly speaking, a mid-plane bias of 20~pc could lead our method to mistake $\Phi(400~\pc)$ for $\Phi(380~\pc)$ or $\Phi(420~\pc)$. It seems possible that our assumption about a perfectly flat Galactic plane might bias our inferred potential of the order of ten per cent, especially for the more distant data samples and for potential values close to the Galactic mid-plane. This was further tested on a three-dimensional simulation in \citetalias{PaperIV}, were we found that the induced bias was contained within the numerical estimate given above.

We have restricted ourselves to parallax uncertainties smaller than $0.05~\mas$. The parallax bias that we correct for is typically of the order of $0.02~\mas$, with potential uncontrolled systematic errors of similar magnitude \citep{2021A&A...649A...2L,edr3_parallax_offset}. For the most distant data samples at $\sqrt{X^2+Y^2} \simeq 3~\kpc$, the relative statistical and systematic uncertainties in parallax are of the order of ten per cent. For our method of inference, a biased parallax propagates into a systematic bias in both height and vertical velocity, which are affected with roughly the same numerical constant. In terms of the inferred gravitational potential, the systematic errors in height and vertical velocity counteract each other. As such, a relative bias in parallax of ten per cent should translate to a smaller relative bias in the inferred gravitational potential of at most a few per cent.

\begin{figure}
	\includegraphics[width=1.\columnwidth]{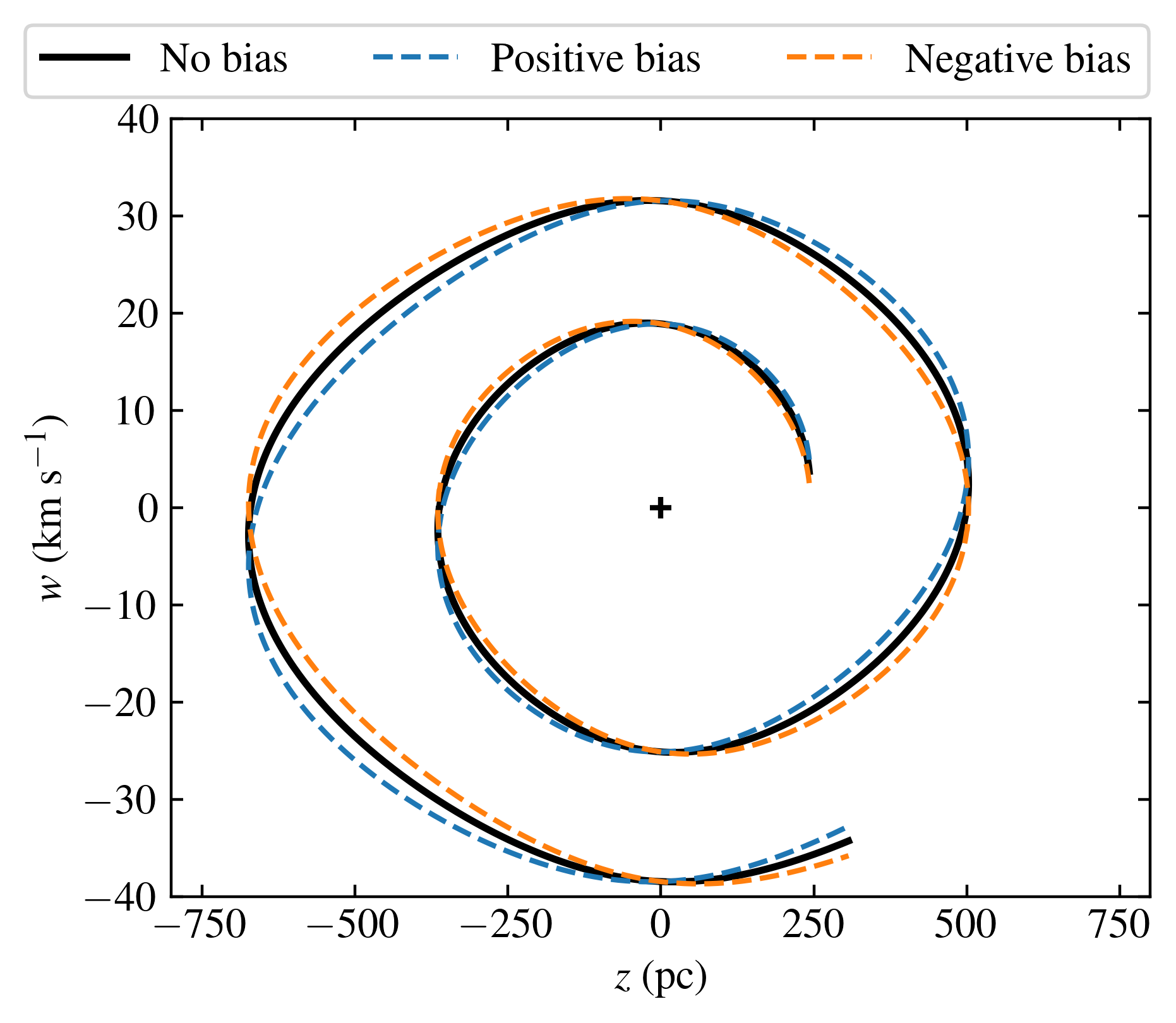}
    \caption{Shape of the phase-space spiral, represented as a one-dimensional line in the $(z,w)$-plane, when biased by faulty radial velocity assignments. The dashed blue and orange lines show the spiral when radial velocity assignment is biased by $\pm 10~\kmsec$, assuming that the spiral is observed at a distance of $\sqrt{X^2+Y^2}=2~\kpc$. The spiral is plotted for vertical energies in range $E_z \in [\Phi(250~\pc), \Phi(800~\pc)]$. The plus sign marks the origin of the $(z,w)$-plane.}
    \label{fig:spiral_RV_bias}
\end{figure}

For stars with missing radial velocities, we assign those values according to the procedure presented in Sect.~\ref{sec:RV_assignment}. This assignment is of course not perfect, which unavoidably introduces an error in the resulting vertical velocities. To the extent that these vertical velocity errors are distributed in a symmetric and unbiased manner, they only serve to soften the resolution of the observed phase-space spiral. However, the distribution of radial velocities is actually somewhat skewed and our estimate of its mean is possibly slightly biased, which propagates into a bias in the stars' vertical velocities ($W$), according to
\begin{equation}\label{eq:W_from_vRV_bias}
    \text{bias}(w) = \text{bias}(v_{\text{RV}}) \times \frac{Z}{\sqrt{X^2+Y^2+Z^2}}.
\end{equation}
This bias in the radial velocity assignment should be sub-dominant with respect to the total dispersion in radial velocity, which is $\sigma_{\text{RV}} \simeq 30~\kmsec$ (varies somewhat between data samples). We make the conservative (i.e. large) estimate that $\text{bias}(v_{\text{RV}}) \lesssim 10~\kmsec$. How this affects the shape of the observed phase-space spiral is illustrated in Fig.~\ref{fig:spiral_RV_bias}, where we have assumed a distance of $\sqrt{X^2+Y^2}=2~\kpc$ and $Z_\odot = 0~\pc$ for simplicity. This spiral inhabits a gravitational potential that follows Eq.~\eqref{eq:phi} with parameter values $\rho_h = \{0.06, 0.03, 0, 0\}~\Msunppcc$, which is representative of the actual data samples and observed spirals that are analysed in this work. We show the shape of the spiral as seen when the radial velocity assignment is positively and negatively biased by $10~\kmsec$. This bias can of course have a more complicated behaviour, for example vary as a function of $z$, but should in either case be roughly constrained to lie within the dashed lines of in Fig.~\ref{fig:spiral_RV_bias}. In the mid-plane, no bias is propagated from the radial velocity assignment, so the three spirals all cross through the same points along the vertical axis defined by $z=0~\pc$. Along the horizontal axis, defined by $w=0~\kmsec$, the three spirals cross through roughly the same points as well. The main difference between the three cases is along the diagonals of the $(z,w)$-plane, where the distance to the origin of the $(z,w)$-plane is biased by around five per cent. While a lot of information about the gravitational potential can be gathered from looking at where the spiral crosses the horizontal and vertical axes, our method uses the full shape of the phase-space spiral in the $(z,w)$-plane and can therefore be biased (although probably most severely in terms of the precise shape of the inferred gravitational potential and matter density distribution, rather than e.g. $\Phi(400~\pc)$). The relative bias in the inferred gravitational potential, as propagated from a biased assignment of radial velocities, should be well contained within
\begin{equation}
    \frac{\text{bias}(\Phi)}{\Phi} \lesssim 0.05 \times \frac{\sqrt{X^2+Y^2}}{2~\kpc}.
\end{equation}

In summary, there are potential significant biases coming from the approximation of a flat Galactic plane, from parallax measurements, and from the assignment of radial velocities. There is also a trade-off, in the sense that the former two types of bias are likely most severe at greater distances, while the latter type of bias is most severe at smaller distances. All in all, systematic biases in terms of $\Phi(400~\pc)$ should be contained to a relative error smaller than about ten per cent, where the more severe biases apply to the most nearby and most distant data samples.

\section{Model of inference}\label{sec:model}

The method of inference used in this work builds upon \citetalias{PaperI} and \citetalias{PaperII}; it more closely resembles the latter, which was also on analysis on \emph{Gaia} EDR3 data. The most significant modification with respect to previous papers is that we include a mask function that depends on $z$, which models the severe selection effects that are present in distant regions of the Galactic disk. Secondly, the gravitational potential is modelled as a sum of four components with scale heights of $\{200,400,800,1600\}~\pc$ (see Eq.~\ref{eq:phi}), which is twice that of previous papers; due to strong selection effects and a potentially warping disk mid-plane, we did not expect to robustly infer the precise shape of the gravitational potential, especially at low heights, and thus opted for a less flexible functional form. A third important modification is that $Z_\odot$ is a fixed parameter, while $W_\odot$ is free. These parameters were free to vary in \citetalias{PaperI}, but fixed parameters in \citetalias{PaperII}. The method of inference used in this work is described below, with an emphasis on its differences with respect to the previous articles in this series, especially the extinction mask function.

\subsection{Phase-space densities}\label{sec:bulk_and_spiral}

The full phase-space density of our model of inference is equal to
\begin{equation}\label{eq:total_density}
\begin{split}
    f(z,w\,|\,\popp) & = B(z,w\,|\,\popp_\text{bulk}) \times \Xi(z \, | \, \popp_{z\text{-mask}}) \\
    & \times \Big[ 1 + m(z,w)\, S(z,w\,|\,\popp_\text{spiral}) \Big].
\end{split}
\end{equation}
It consists of three distributions that are free to vary: a bulk density distribution, $B(z,w\,|\,\popp_\text{bulk})$; an extinction mask in height, $\Xi(z \, | \, \popp_{z\text{-mask}})$; and a relative spiral density perturbation, $S(z,w\,|\,\popp_\text{spiral})$. These three distributions depend on the model's free parameters $\popp = \{\popp_\text{bulk},\, \popp_{z\text{-mask}},\, \popp_\text{spiral}\}$, which are listed in Table~\ref{tab:model_parameters}. The quantity $m(z,w)$ is a fixed inner mask function. All these distributions are defined below.

{\renewcommand{\arraystretch}{1.6}
\begin{table}[ht]
	\centering
	\caption{Free parameters in our model of inference.}
	\label{tab:model_parameters}
    \begin{tabular}{| l | l |}
		\hline
		$\popp_\text{bulk}$  & Bulk phase-space density parameters \\
		\hline
		$a_k$ & Weights of the Gaussian mixture model \\
		$\sigma_{z,k}$, $\sigma_{w,k}$ & Dispersions of Gaussian mixture model \\
		$W_\odot$ & Vertical velocity of the Sun relative to the bulk \\
		\hline
		\hline
		$\popp_{z-\text{mask}}$  & Mask in $z$ \\
		\hline
		$\hat{a}_l$ & Amplitudes \\
		$\hat{z}_l$ & Means \\
		$\hat{\sigma}_{z,l}$ & Dispersions \\
		\hline
		\hline
		$\popp_\text{spiral}$  & Spiral phase-space density parameters \\
		\hline
		$\rho_{h=\{1,2,3,4\}}$ & Mid-plane matter densities \\
		$t$ & Time since the perturbation was produced \\
		$\tilde{\varphi}_0$ & Initial angle of the perturbation \\
		$\alpha$ & Relative density amplitude of the spiral \\
		\hline
	\end{tabular}
\end{table}}

The bulk density distribution is equal to
\begin{equation}\label{eq:bulk_density}
    B(z,w\,|\,\popp_\text{bulk}) =
    \sum_{k=1}^{K} a_k \,
    \dfrac{\exp\Bigg(-\dfrac{z^2}{2\sigma_{z,k}^2}\Bigg)}{\sqrt{2\pi\sigma_{z,k}^2}} \,
    \dfrac{\exp\Bigg[-\dfrac{(W+W_\odot)^2}{2\sigma_{w,k}^2}\Bigg]}{\sqrt{2\pi\sigma_{w,k}^2}},
\end{equation}
which is a Gaussian mixture model, with the constraints that all Gaussians are centred on the same point in the $(z,w)$-plane and have zero correlations between the $z$ and $w$ directions. The respective Gaussians are labelled by an index $k=\{1,2,...,K\}$. In traditional methods that are based on the assumption of a steady state, the bulk density is used to infer the gravitational potential, for example by fitting it to data under the requirement that it fulfils the stationary collisionless Boltzmann equation. In our modelling, the bulk density does not inform, nor is informed by, the gravitational potential; as such, it is solely a background distribution fitted in order to extract the shape of the spiral.

The data samples that are studied in this work suffer from severe selection effects, mainly due to stellar crowding and dust extinction. Especially because of dust, these selection effects are difficult to estimate. While incompleteness is to a significant extent induced by our cuts in parallax uncertainty (see Sect.~\ref{sec:data_cuts}), it is not enough to model the incompleteness as a function of apparent magnitude and position on the sky. Dust extinction for disk stars at these distances can be several magnitudes in the \emph{Gaia} \emph{G}-band, and in the relevant spatial volumes current three-dimensional dust maps are not very precise (e.g. figure 9 in \citealt{2019A&A...625A.135L}).
Thankfully, due to the nature of this selection, we can assume that it depends mainly on spatial position, rather than velocity, and that it is most severe close to the Galactic mid-plane. Because of this, we modelled the selection effects by using an extinction mask function written as
\begin{equation}
	\Xi(z \, | \, \popp_{z\text{-mask}}) = 1 - \sum_{l=1}^{L} \hat{a}_l 
	\exp\Bigg[-\dfrac{(z-\hat{z}_l)^2}{2\hat{\sigma}_{z,l}^2}\Bigg].
\end{equation}
This mask function is fitted to data and its free parameters are all written with hats and with an index $l=\{1,2,...,L\}$. The inclusion of this mask function constitutes the most significant modification with respect to the method used in \citetalias{PaperI} and \citetalias{PaperII}. The purpose of this mask function is not to model selection effects very accurately; in fact, we expect there to be degeneracies between the fitted extinction mask and bulk density distribution. Rather, the purpose of this mask is to facilitate the extraction of the phase-space spiral's shape in the $(z,w)$-plane. This aspect of our method has been thoroughly tested in \citetalias{PaperIV}, where we subjected the data to selection effects similar to what is seen in this work. In these tests, could accurately extract the spiral and vertical gravitational potential, despite degeneracies between the fitted bulk and extinction mask. This point is further supported by the results of \citetalias{PaperII}, where the respective data samples were inconsistent in terms of the scale heights of the fitted bulk distributions (due to the strong distance dependence of the \emph{Gaia} radial velocity sample) but consistent in terms of their extracted spirals and vertical gravitational potentials.

The spiral relative density perturbation is identical to how it was defined in \citetalias{PaperII}. It is written as
\begin{equation}\label{eq:spiral_rel_density}
    S(z,w\,|\,\popp_\text{spiral}) =
    \alpha \cos\Big[ \varphi(z,w\,|\, \rho_h)-\tilde{\varphi}(t,E_z \,|\, \rho_h,\tilde{\varphi}_0) \Big].
\end{equation}
This depends on the angles of a given phase-space point, which is defined as
\begin{equation}\label{eq:angle_of_z}
\begin{split}
& \varphi(z,w \,|\, \rho_h) = \\
& \begin{cases}
    2 \pi P^{-1}{\displaystyle\int_0^{|z|}} \dfrac{\de z'}{\sqrt{2[E_z-\Phi(z' \,|\, \rho_h)]}} & \text{if}\,z\geq0\,\text{and}\,w\geq0, \\
    \pi - 2 \pi P^{-1}{\displaystyle\int_0^{|z|}} \dfrac{\de z'}{\sqrt{2[E_z-\Phi(z' \,|\, \rho_h)]}} & \text{if}\,z\geq0\,\text{and}\,w<0, \\
    \pi + 2 \pi P^{-1}{\displaystyle\int_0^{|z|}} \dfrac{\de z'}{\sqrt{2[E_z-\Phi(z' \,|\, \rho_h)]}} & \text{if}\,z<0\,\text{and}\,w<0, \\
    2\pi - 2 \pi P^{-1}{\displaystyle\int_0^{|z|}} \dfrac{\de z'}{\sqrt{2[E_z-\Phi(z' \,|\, \rho_h)]}} & \text{if}\,z<0\,\text{and}\,w\geq0,
\end{cases}
\end{split}
\end{equation}
as well as that of the spiral,
\begin{equation}\label{eq:angle_of_time}
    \tilde{\varphi}(t,E_z \,|\, \rho_h,\tilde{\varphi}_0) = \tilde{\varphi}_0 + 2\pi\frac{t}{P(E_z \,|\, \rho_h)}.
\end{equation}
In these expressions, $z_\text{max}$ is the maximum height that a star reaches, and $P$ is the period of vertical oscillation, given by
\begin{equation}\label{eq:period}
    P(E_z\,|\, \rho_h) = \oint \frac{\de z}{w} =  4\int_0^{z_\text{max}} \frac{\de z}{\sqrt{2[E_z-\Phi(z \,|\, \rho_h)]}}.
\end{equation}

The quantity $m(z,w)$ is a inner mask function. It defines the inner boundary around the origin of the $(z,w)$-plane, within which the phase-space spiral is not seen. The mask function is equal to
\begin{equation}
    m(z,w) = \sigm \Bigg\{ 10 \times \Bigg[
    \frac{z^2}{(300~\pc)^2} + \frac{w^2}{\text{Std}^2(w)} -1
    \Bigg] \Bigg\} ,
\end{equation}
where $\text{Std}(w)$ is the data sample's standard deviation in vertical velocity, and
\begin{equation}\label{eq:sigmoid}
    \sigm(x) = \frac{1}{1+\exp(-x)}
\end{equation} 
is a sigmoid function. This differs slightly from how our method was formulated in \citetalias{PaperI} and \citetalias{PaperII}, where the inner mask function was defined in terms of a boundary in vertical energy equal to $\Phi(300~\pc)$; in those earlier paper, the inner mask's boundary in velocity varied with the free parameters $\rho_h$. We opted to change this in order to make our algorithm somewhat more stable; in this updated version, the region of the $(z,w)$-plane that the fitted spiral is sensitive to does not change with the free parameters of our model.

\subsection{Data likelihood and masks}\label{sec:likelihood_and_masks}

The fitting procedure is essentially the same is in the previous papers in this series. We fitted the bulk density and mask function ($\popp_\text{bulk}$ and $\popp_{z\text{-mask}}$) in a first step, without any spiral density perturbation present. In a second step, we fitted the relative phase-space density of the spiral ($\popp_\text{spiral}$) while the bulk and mass function remain fixed. The data is reduced to a histogram in the $(z,w)$-plane, written $\data_{i,j}$, where the respective bins are labelled by $(i,j)$ and have widths of $20~\pc$ and roughly $1~\kmsec$. The logarithm of the likelihood is written as
\begin{equation}\label{eq:likelihood}
\begin{split}
    \ln\, \mathcal{L}(\data_{i,j}\,|\,\popp) = &
    - \sum_{i,j} M(z_i,w_i) \times \dfrac{[\data_{i,j}-f(z_i,w_i\,|\,\popp)]^2}{2 f(z_i,w_i\,|\,\popp)} \\
    & + \{\text{constant term}\},
\end{split}
\end{equation}
where
\begin{equation}\label{eq:mask}
    M(z,w) =
    \text{sigm} \Bigg[ -10\, \Bigg( \frac{z^2}{z_\text{lim.}^2} + \frac{w^2}{w_\text{lim.}^2} - 1 \Bigg) \Bigg],
\end{equation}
is an outer mask function, which defines the circular region in the $(z,w)$-plane in which we perform our fit. The two steps of our fitting procedure use different outer boundaries: for the first step, when fitting the bulk density, we set $z_\text{lim.} = 7/2 \times 300~\pc$ and $w_\text{lim.} = 7/2 \times \text{Std}(w)$; for the second step, when fitting the spiral, we set $z_\text{lim.} = 7/3 \times 300~\pc$ and $w_\text{lim.} = 7/3 \times \text{Std}(w)$. This is similar to the outer bounds that we applied in previous papers in this series, although the boundary in vertical velocity varies between data samples, mainly because $\text{Std}(w)$ becomes smaller with greater Galactocentric radius.

In order to make the minimisation algorithm computationally tractable, it is implemented in \textsc{TensorFlow}. Readers can refer to \citetalias{PaperI} for further details.

\section{Results}\label{sec:results}

When running our minimisation algorithm, we used a total number of $K=6$ Gaussians for the bulk density distribution and $L=6$ Gaussians for the mask function. We found that modelling of the selection effects did not improve when increasing $L$.

\begin{figure}
	\includegraphics[width=1.\columnwidth]{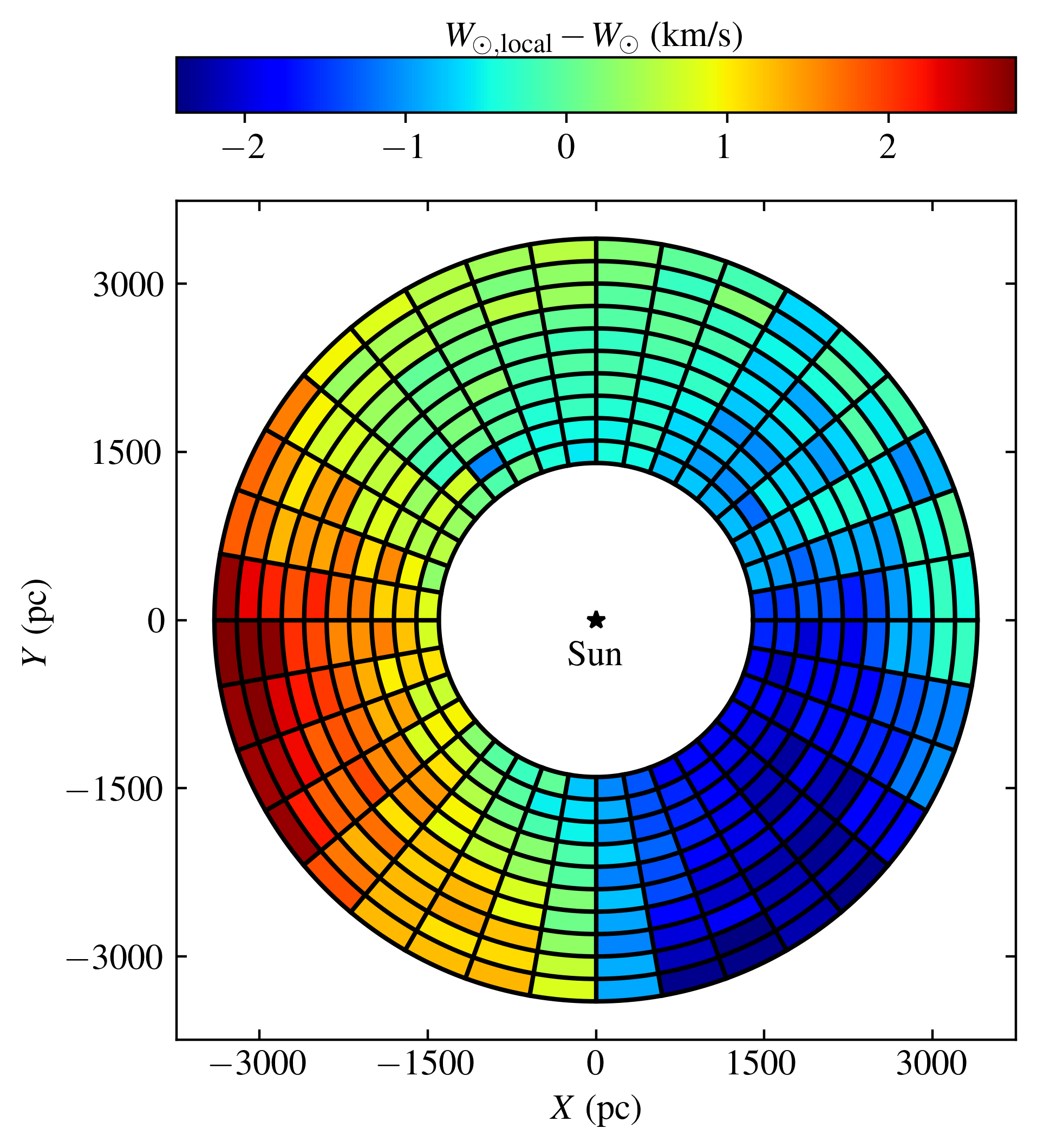}
    \caption{Bulk vertical velocity relative to that of the immediate solar neighbourhood. This is equivalent to $W_{\odot,\text{local}}-W_\odot$, where $W_{\odot,\text{local}}=7.25~\kmsec$ and $W_\odot$ is a free parameter fitted in the first step of our minimisation algorithm. The Galactic centre is located to the right, while the direction of Galactic rotation is upwards.}
    \label{fig:W_corrections}
\end{figure}

The mode of the vertical velocity distribution, which was inferred in the first step of our minimisation algorithm, is shown in Fig.~\ref{fig:W_corrections}. It is expressed in terms of $W_{\odot,\text{local}} - W_\odot$, where $W_{\odot,\text{local}} = 7.25~\kmsec$ is the vertical velocity of the Sun with respect to the Galactic disk in the immediate solar neighbourhood, while $W_\odot$ is the corresponding free parameter for the Galactic disk at the position of the respective data samples. There is significant dependence on $X$ and $Y$ with regards to this parameter, with especially low values in the direction of $l\simeq 320~\deg$ and high values in the direction of $l\simeq 190~\deg$. These result agree well with those found by \citet[][see their figure 10]{gaia_kinematics}, \citet[][see their figure 3]{2018MNRAS.481L..21P}, and \citet[][see their figure 6]{2021arXiv210904696M}.

\begin{figure*}
	\includegraphics[width=1.\textwidth]{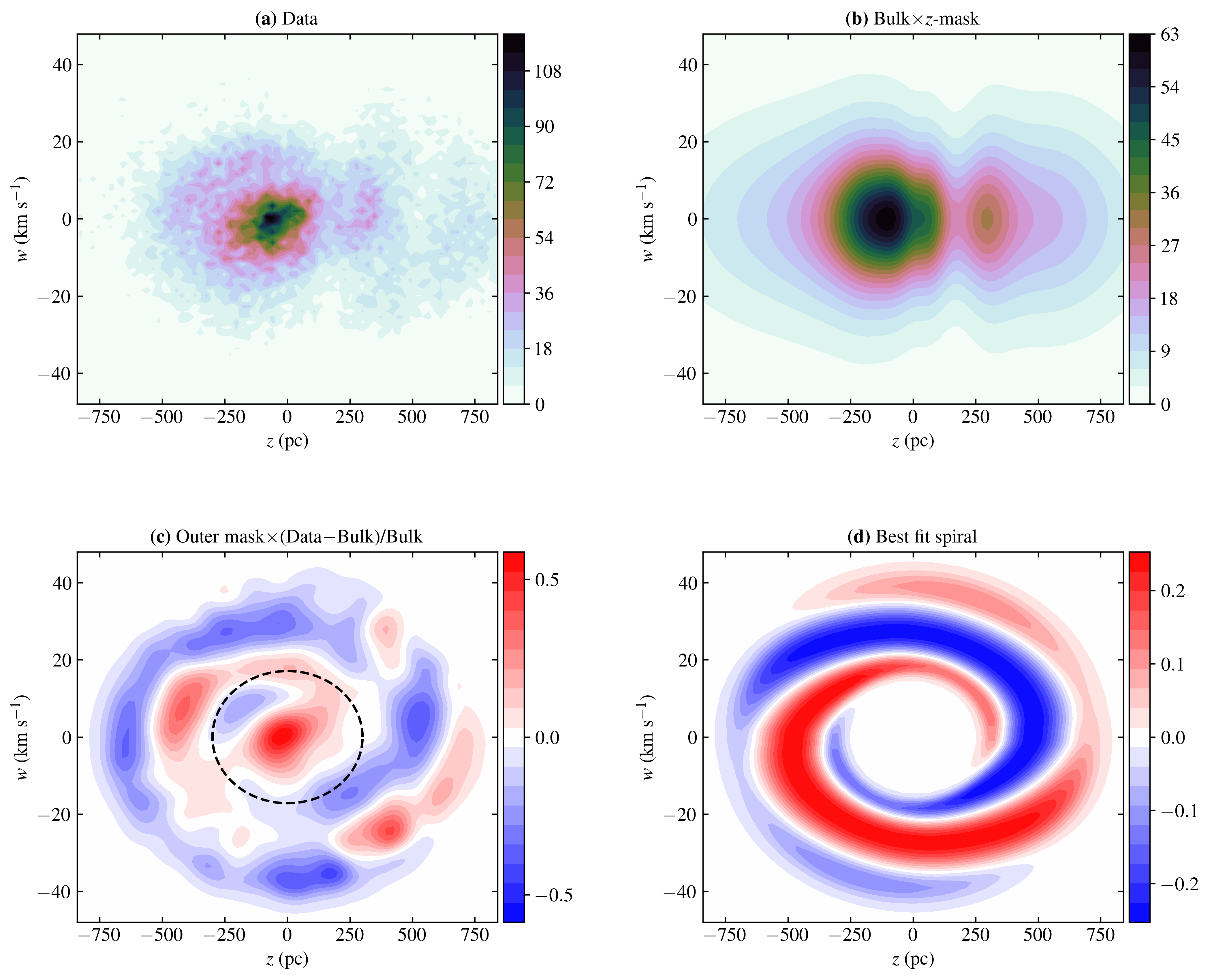}
    \caption{Data and fitted phase-space density of the data sample with $l \in [120,130]~\deg$ and $\sqrt{X^2+Y^2} \in [2000,2200]~\pc$. All four panels span the same range of the $(z,w)$-plane and show:
    {\bf (a)} the data histogram;
    {\bf (b)} the fitted bulk and $z$-mask;
    {\bf (c)} the spiral as seen in the data after removing the bulk and $z$-mask, where the dashed black line shows the boundary of the inner mask function;
    {\bf (d)} the relative phase-space density perturbation of the best fit spiral.}
    \label{fig:spiral_illustrative_example}
\end{figure*}

In Fig.~\ref{fig:spiral_illustrative_example}, we show the data histogram, fitted bulk density, and spiral as seen in the data and best fit, for the data sample with $l \in [120,130]~\deg$ and $\sqrt{X^2+Y^2} \in [2000,2200]~\pc$. This specific data sample was chosen as a representative and illustrative example of how our algorithm can extract and fit the phase-space spiral despite severe selection effects. Looking at panels {\bf (a)} and {\bf (b)}, which show the data histogram and the fitted bulk times $z$-mask, there are clear extinction features in the form of vertical stripes, seen most clearly at the height of $z\simeq 200~\pc$. In panel {\bf (a)}, it is very difficult to see the shape of the spiral by eye. However, a spiral clearly emerges in the data after removing the bulk times $z$-mask in panel {\bf (c)}, and its shape is well reproduced by the fitted spiral in panel {\bf (d)}. The asymmetry of these extinction features with respect to the Galactic mid-plane illustrate why $Z_\odot$ is not a free parameter in our model, as this quantity would be strongly biases by selection effects. 

\subsection{Disqualified and dubious data samples}
\label{sec:disqualified_samples}

After running our minimisation algorithm, we inspected the results of the 360 data samples by eye. Many data samples had to be removed altogether or be marked as dubious, due to severe selection effects or some other systematic issue.

The most significant reason for disqualifying data samples is that of selection effects. This is most severe in the approximate direction of the Galactic centre, where we had to disqualify the region where $|l|\lesssim 50~\deg$ altogether. For the majority of these disqualified data samples, no convincing spiral could be seen by eye after fitting the first step of our minimisation procedure (i.e. in plots corresponding to panel {\bf (c)} in Fig.~\ref{fig:spiral_illustrative_example}). We also disqualified a few data samples were the phase-space spiral could be seen by eye, but our algorithm did not manage to fit that spiral correctly, despite being run several times with different initial conditions. For some cases the fitted spiral had some qualitative differences with respect to the spiral seen by eye, in which case we marked them as dubious. Additionally, in some cases our fit agreed reasonably well with a spiral that was somewhat vaguely discernible by eye, in which case we marked those data samples as dubious.

\begin{figure*}
    \centering
	\includegraphics[width=.8\textwidth]{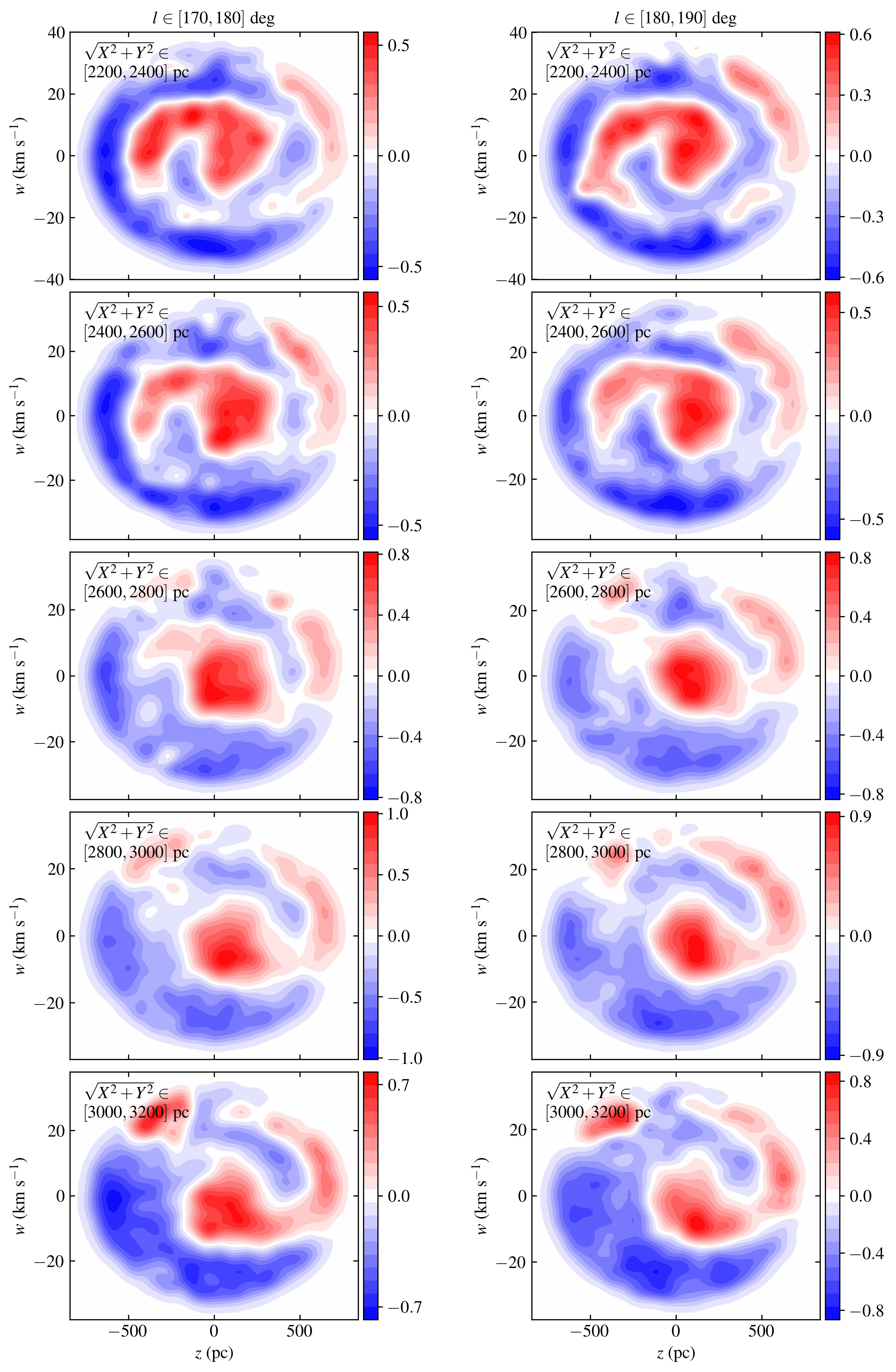}
    \caption{Phase-space spiral seen in the direction of the Galactic anti-centre, for data samples with $l \in [170,180]$ degrees (left column) and $l \in [180,190]$ degrees (right column), and $\sqrt{X^2+Y^2}$ in range 2200--3200 pc (increasing from top to bottom). The respective panels show the relative difference between the data histogram and fitted bulk density distribution: $M\times(\data-B)/B$. The horizontal axis is identical for all panels.}
    \label{fig:spiral_anticentre}
\end{figure*}

The direction of the Galactic anti-centre is less plagued by selection, but suffers from another type of systematic issue. At a distance of about $2600~\pc$, the phase-space spiral undergoes a fairly dramatic change in shape over quite small distance scales. This is illustrated in Fig.
\ref{fig:spiral_anticentre}, showing the phase-space spiral seen in the data after removing the bulk and $z$-mask (corresponding to panel {\bf (c)} in Fig.~\ref{fig:spiral_illustrative_example}), for data samples within the spatial volume defined by $l\in [170,190]~\deg$ and $\sqrt{X^2+Y^2}\in [2200,3000]~\pc$. If we compare the closest and most distant data samples in Fig.
\ref{fig:spiral_anticentre}, the arm of the phase-space spiral protrudes from either the left or right. It seems possible that this is due to poor data accuracy, as an effect of the kinematics at low heights being blurred. If this is indeed something physical, we expect to find out with a more dedicated analysis using future \emph{Gaia} data releases. Either way, we did not account for the spiral to have any dependence on the spatial direction parallel to the Galactic plane; this can only be a reasonable approximation if the shape of the phase-space spiral is fairly constant between neighbouring data samples. For this reason, we either disqualified or marked the data sample as dubious in this spatial region.

In summary, we disqualified a total number of 181 data samples, and marked 85 data samples as dubious. That left us with a total number of 94 well behaved data samples, referred to as good data samples below. We show a few examples of dubious data samples in Appendix~\ref{app:examples}.

\subsection{The inferred gravitational potential}
\label{sec:inferred_grav_pot}

\begin{figure}
	\includegraphics[width=1.\columnwidth]{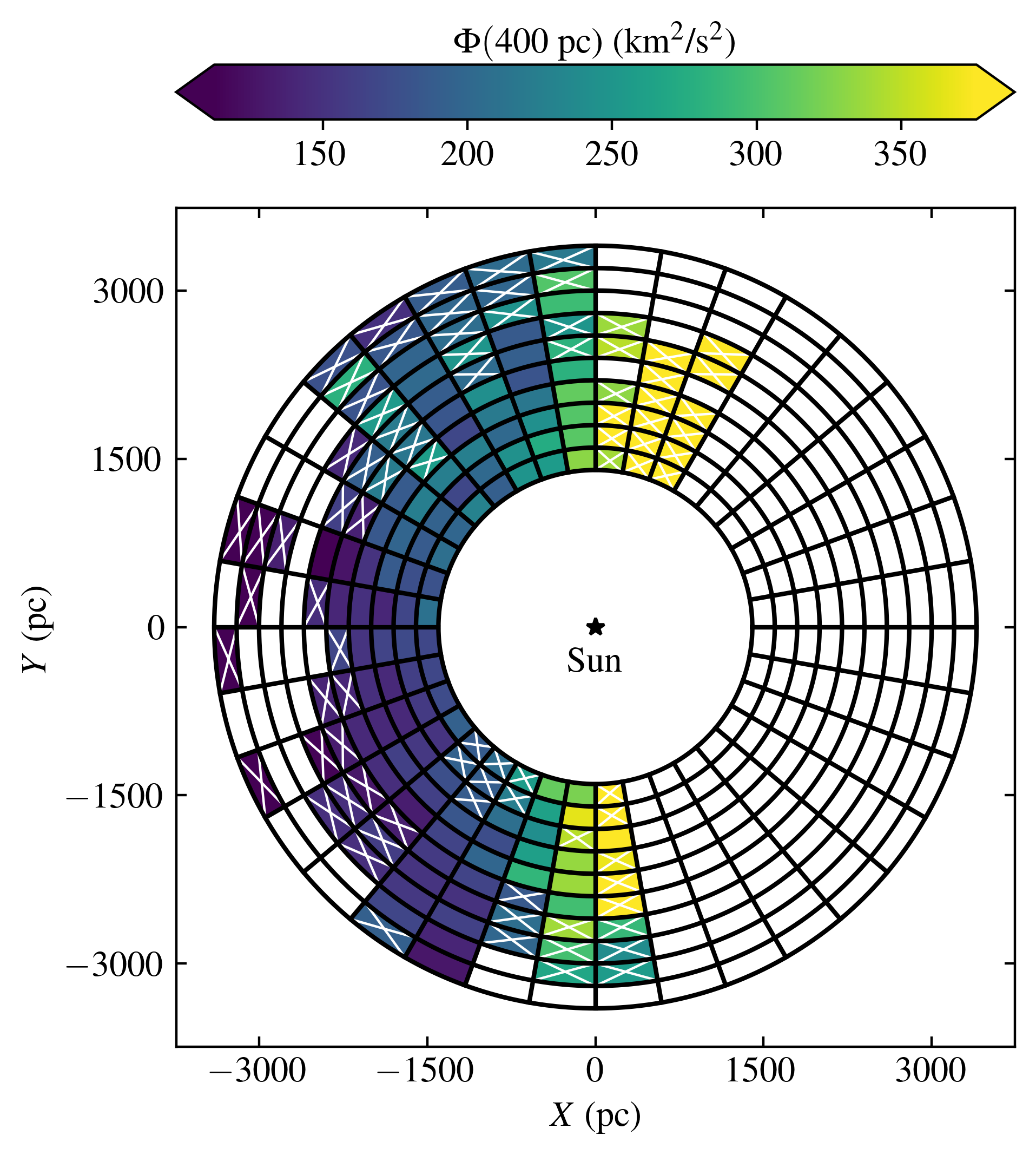}
    \caption{Inferred gravitational potential at a height of $|z|=400~\pc$, for the respective data samples. Disqualified data samples are left blank and dubious data samples are marked with a white cross.}
    \label{fig:inferred_phi_400pc}
\end{figure}

\begin{figure}
	\includegraphics[width=1.\columnwidth]{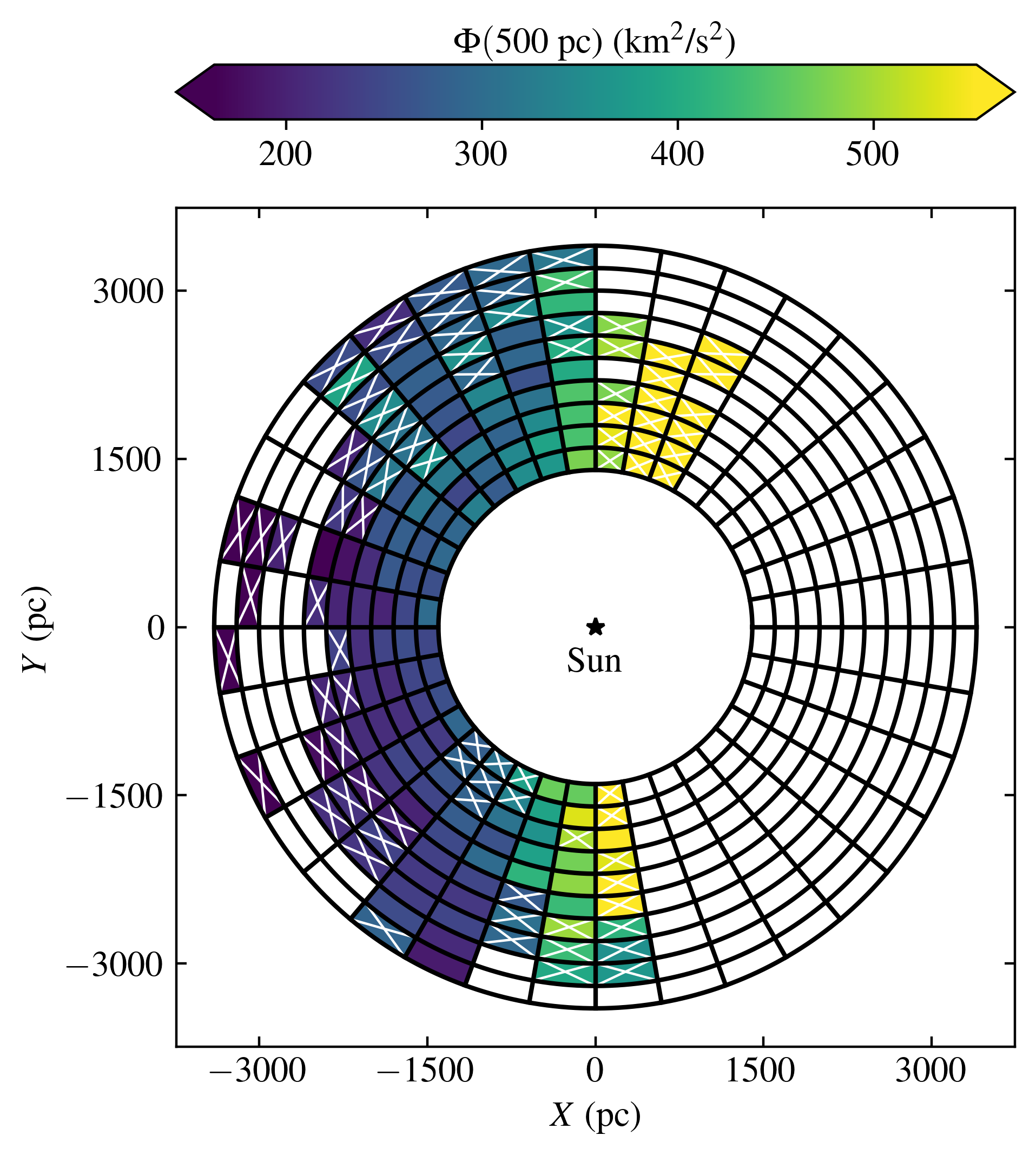}
    \caption{Same as Fig.~\ref{fig:inferred_phi_400pc}, but for $|z|=500~\pc$.}
    \label{fig:inferred_phi_500pc}
\end{figure}

We present our main results for the vertical gravitational potential in terms of $\Phi(400~\pc)$ and $\Phi(500~\pc)$, because the potential in this height range was found to be the most robustly inferred quantity in our tests on simulations in \citetalias{PaperI}. In Figs.~\ref{fig:inferred_phi_400pc} and \ref{fig:inferred_phi_500pc}, we show the inferred gravitational potential values for our good and dubious data samples, at their respective positions in the $(X,Y)$-plane. This map is more or less split in half in terms of what data samples produced useful results, where the direction of the Galactic centre is left completely blank. Overall, the dubious data samples agree quite well with the general spatial dependence of the good data samples. As expected, there is a clear trend of lower values with greater Galactocentric radius (i.e. in the direction of negative $X$). We also see some variations as a function of azimuth, where the direction of $l\simeq 225~\deg$ (bottom left quadrant) has somewhat lower values compared to $l\simeq 135~\deg$ (top left quadrant); this is discussed further below and in Sect.~\ref{sec:discussion}. We also show results for other heights in Appendix~\ref{app:suppl_plots}, as well as the inferred vertical acceleration and the inferred time since the perturbation was produced (the model parameter $t$).

The maps shown in Figs.~\ref{fig:inferred_phi_400pc} and \ref{fig:inferred_phi_500pc} are fairly smooth. If we compare neighbouring good data samples (in both spatial directions), the relative difference in $\Phi(500~\pc)$ between them has a median and mean value of 10\% and 13\%, respectively (where the latter is significantly higher due to a few strong outliers). This could very well be explained, at least in part, by intrinsic variability in the vertical gravitational potential between the spatial locations of the respective data samples. Perhaps most importantly, the cold gas present in the Galactic disk, constituting roughly one fourth of the thin disk surface density \citep{2015ApJ...814...13M}, is highly structured on smaller spatial scales. The dust distribution, which traces the most significant gas component of cold atomic gas, has significant structure on scales of around 100 pc \citep{2003A&A...411..447L,2019A&A...625A.135L}, as does the second-most significant component of cold molecular gas \citep{2001ApJ...547..792D}. Considering that our method has a statistical accuracy of a few per cent at best, which is what we had when we applied it to simulations in \citetalias{PaperI}, the difference in gravitational potential values between neighbouring data samples is reasonable, at the very least when discounting the few strong outliers.

\begin{figure}
	\includegraphics[width=1.\columnwidth]{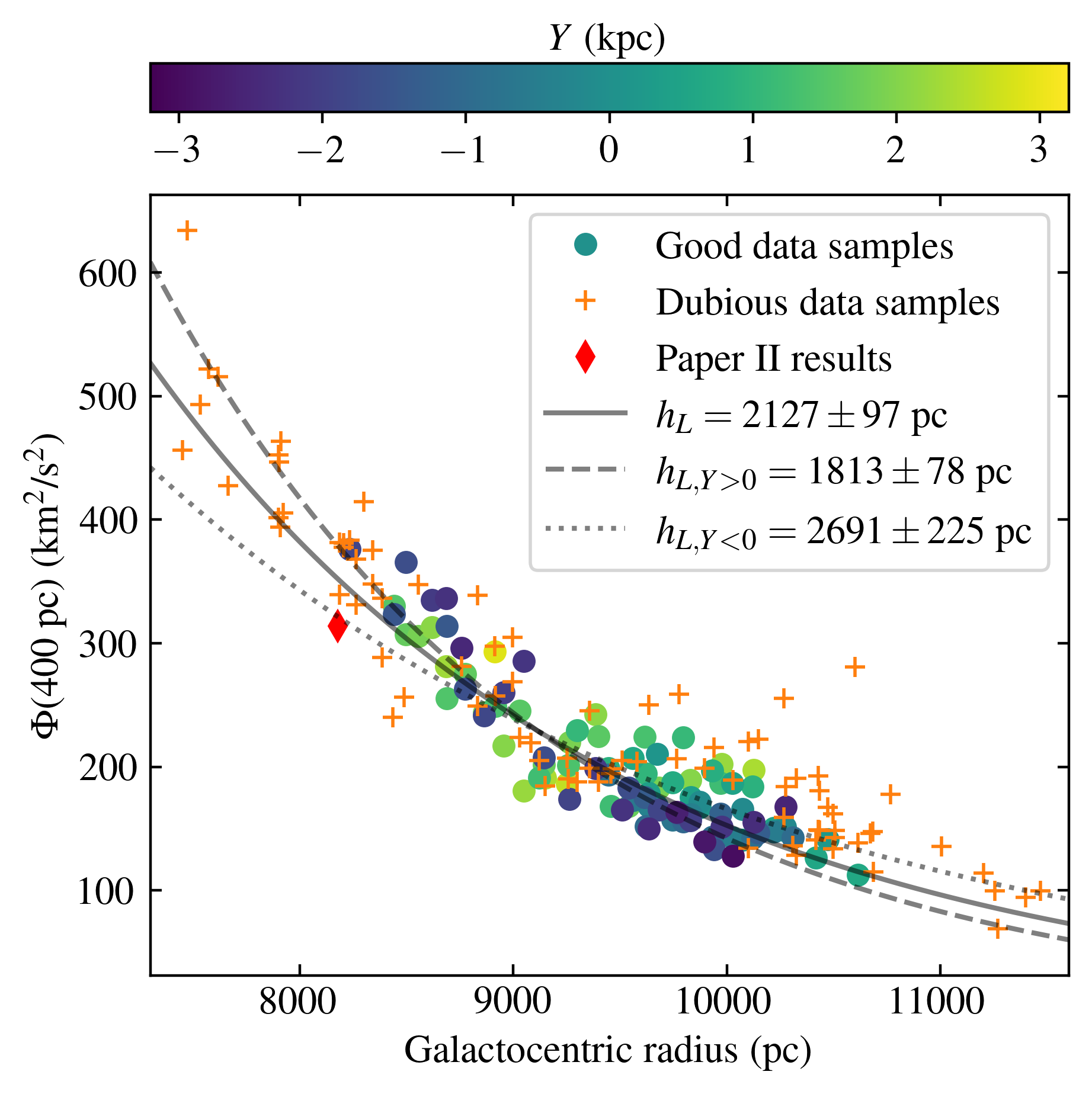}
    \caption{Inferred gravitational potential at a height of $|z|=400~\pc$, as a function of Galactocentric radius. The results of the good data samples are marked with circles, which are colour coded according to the data samples' respective $Y$ coordinate. The dashed line shows a best fit exponential curve with respect to the good data samples. We also show the results coming from the dubious data samples, and the result of \citetalias{PaperII} for the immediate solar neighbourhood.}
    \label{fig:phi_of_R_400pc}
\end{figure}

\begin{figure}
	\includegraphics[width=1.\columnwidth]{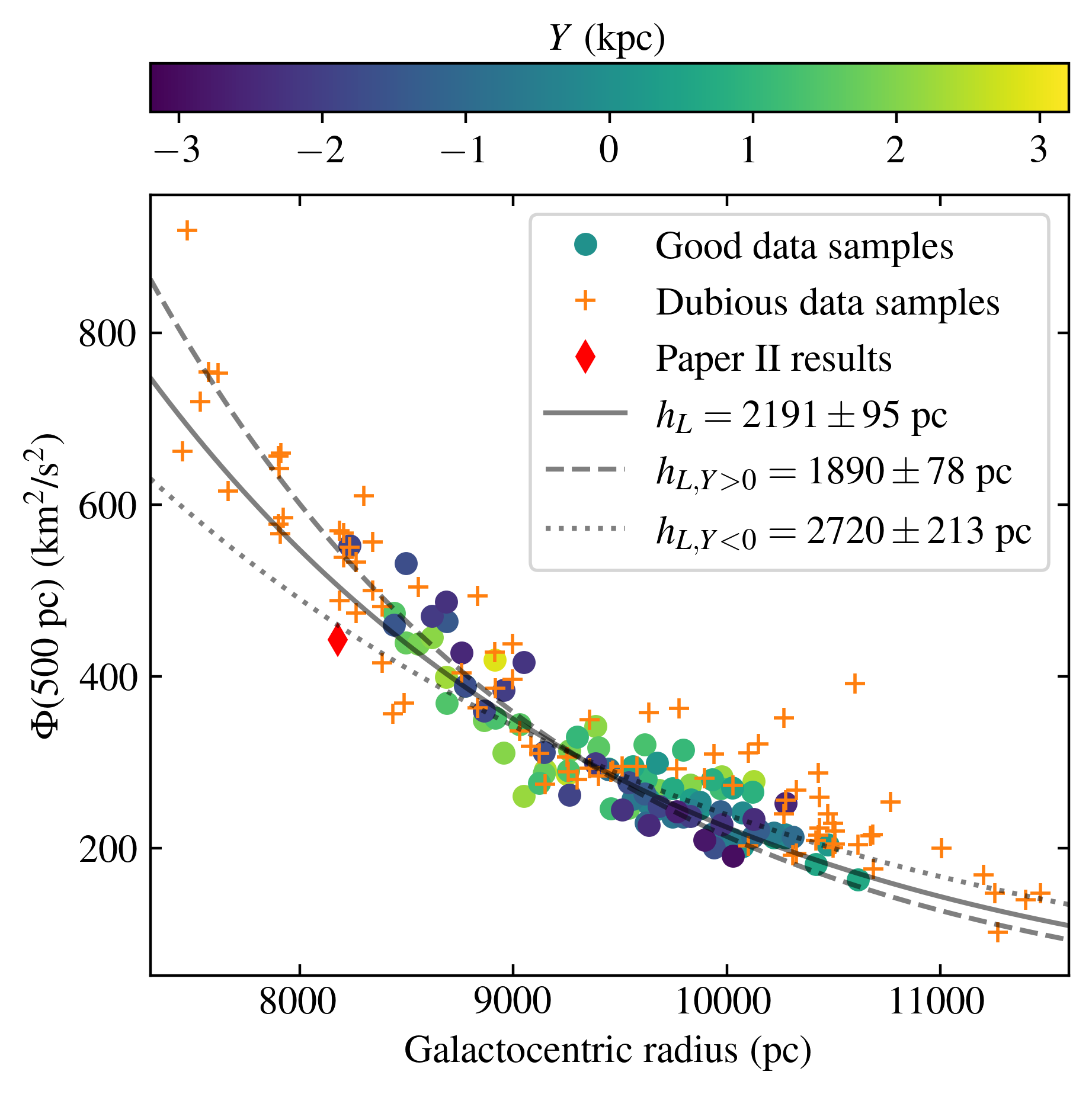}
    \caption{Same as Fig.~\ref{fig:phi_of_R_400pc}, but for $|z|=500~\pc$.}
    \label{fig:phi_of_R_500pc}
\end{figure}

The dependence of $\Phi(400~\pc)$ and $\Phi(500~\pc)$ with respect to Galactocentric radius is visible in Figs.~\ref{fig:phi_of_R_400pc} and \ref{fig:phi_of_R_500pc}. The circular markers, representing good data samples, are colour coded according to their spatial $Y$ coordinate, which highlights the broken axisymmetry seen at higher Galactocentric radii. The difference between these regions is of the order of 10--20~\%. Furthermore, the results for the immediate solar neighbourhood in \citetalias{PaperII}, shown as a diamond marker in Figs.~\ref{fig:phi_of_R_400pc} and \ref{fig:phi_of_R_500pc}, have a smaller value with respect to the data samples of this work that are at a similar Galactocentric radius, with a similar relative difference. These variations are roughly consistent with our 10~\% estimate of possible systematic biases that could be present in our study (see in Sect.~\ref{sec:biases}). When comparing the results presented in Fig.~\ref{fig:phi_of_R_400pc} and \ref{fig:phi_of_R_500pc}, as well as the supplementary plots in Appendix~\ref{app:suppl_plots}, its clear that the variations as a function of azimuth are larger (in a relative sense) for gravitational potential values at lower heights. For this reason, it seems plausible that our results could be biased by a warping of the Galactic disk; this is further discussed in Sect.~\ref{sec:discussion}.

To the good data samples, we fitted an analytic function proportional to $\exp(-R/h_L)$, where $R$ is the Galactocentric radius and $h_L$ is a disk scale length. We also fitted separate curves for the groups of good data samples with either only positive or only negative $Y$-coordinates. These best fitted curves are seen in Figs.~\ref{fig:phi_of_R_400pc} and \ref{fig:phi_of_R_500pc} as grey lines. Their respective inferred scale lengths---labelled $h_L$, $h_{L,Y>0}$, and $h_{L,Y<0}$---are also written out in the figure legends, where the uncertainty comes from assuming a measurement uncertainty for all data samples that makes the scaled $\chi^2$ value equal to unity. Clearly, when choosing different spatial cuts, in this case splitting the spatial volume in half along the solar azimuth, we obtain very discrepant results, regardless of what gravitational potential or vertical acceleration value we consider.

\section{Discussion}\label{sec:discussion}

In this work, we have weighed the Galactic disk for distances in the range 1.4--3.4 kpc, with a high spatial resolution in the directions parallel to the Galactic plane, using the \emph{Gaia} EDR3 proper motion sample. We were able to relax otherwise common assumptions about the Milky Way's large scale gravitational potential and matter density distribution, most importantly the assumptions of axisymmetry and the exponential decay of the disk mass as a function of Galactocentric radius.

Despite using the \emph{Gaia} EDR3 proper motion sample, the data samples we constructed were still subject to severe selection effects, due to the combination of their large spatial distance and our cuts in data quality. For this reason, we expanded our model of inference by adding a simple data-driven extinction function that models incompleteness as a function of height (written $\Xi$, see Sect.~\ref{sec:bulk_and_spiral}). While our modelling of incompleteness effects was far from perfect, it was largely sufficient for extracting the shape of the phase-space spiral, which is what we need in order to infer the gravitational potential. In this manner, our method differs qualitatively from traditional methods that are based on the assumption of a steady state; such methods are highly sensitive to selection and the gravitational potential that they infer can only be as accurate as the underlying completeness model. This fundamental difference and robustness of our method is exemplified in \citetalias{PaperII}, where nearby data samples (100~pc wide in Galactocentric radius) produced stable results despite very varied selection effects (for example, the scale height of the included stars differed by up to 50~\% between data samples). We could only apply our complete method of inference to 179 out of 360 data samples, where the majority of disqualified data samples were in the direction of the Galactic centre, where stellar crowding and dust extinction is most severe. Out of the 179 data samples, an additional 85 were marked as dubious when comparing the observed and fitted spiral by eye (see Sect.~\ref{sec:disqualified_samples} for details).

The first step of our minimisation algorithm, where we fit the bulk density distribution, was applied to all 360 data samples. Thus we could infer the vertical velocity of the Sun with respect to the bulk, written $W_\odot$, for the full region of the Galactic disk, which agreed well with the results of more dedicated analyses (e.g. \citealt{gaia_kinematics,2021arXiv210904696M}). In the studied region, the bulk vertical velocity relative to that of the immediate solar neighbourhood (the quantity shown in Fig.~\ref{fig:W_corrections}) lies in the range $[-2.4,2.8]~\kmsec$. However, the data samples with the most significant outlier values are either disqualified or marked as dubious; for our good data samples, the same quantity lies in the range $[-1.0,1.6]~\kmsec$. As such, it seems like we are constraining ourselves to a spatial region of the Galactic disk with less significant warping.

In terms of the inferred gravitational potential, we present our results in terms of $\Phi(400~\pc)$ and $\Phi(500~\pc)$, which were found to be the most robustly inferred quantities in our tests on simulations in \citetalias{PaperI}. Because it seems like our results are somewhat biased especially at lower heights, we consider $\Phi(500~\pc)$ to be more robust for this specific work. The gravitational potential value is close to proportional to the total surface density of the thin disk; this linear relationship holds to the extent that the general shape of the matter density distribution as a function of height is the same for all relevant data samples (i.e. they differ only in terms of their respective amplitudes). In terms of our inferred results, $\Phi(500~\pc)$ is proportional to the inferred total surface density within $|z|<300~\pc$, to a relative precision of 0.5~\%. Due to this strong linear relation, we use the quantities $\Phi(500~\pc)$ and thin disk surface density more or less interchangeably in the remainder of this paper, as well as in the abstract.

Our results are not perfectly axisymmetric, but have relative variations of the order of 10--20~\% with respect to the azimuth. This is seen most clearly for data samples at the Galactocentric radius of around 10~kpc, but also in comparison with the immediate solar neighbourhood as analysed in \citetalias{PaperII}. This variation is roughly the same order of magnitude as the considered systematic biases, evaluated to roughly 10~\% in Sect.~\ref{sec:biases}. Another interesting facet of our results, seen even more clearly in the supplementary plots of Appendix~\ref{app:suppl_plots}, is that these discrepancies are more severe (in a relative sense) when comparing gravitational potential values at smaller heights. This could be explained by a systematic bias coming from an erroneous assumption of a perfectly flat Galactic plane, which would affect the inferred potential especially at low heights. In summary, the broken axisymmetry that we observe could well be explained by a systematic bias, most likely due to variations in height of the Galactic disk mid-plane. For future analysis similar to this work, we would need to include, and possibly produce ourselves, detailed maps of vertical variations to the Galactic disk mid-plane.

Using the inferred value of $\Phi(500~\pc)$ and assuming an exponential decay as a function of Galactocentric radius, we infer a thin disk scale length of $2.2\pm 0.1~\kpc$. This is in decent agreement with previous studies by for example \citet[][$2.15\pm 0.14~\kpc$]{Bovy:2013raa}, \citet[][$2.6\pm 0.2~\kpc$]{2016ARA&A..54..529B}, and \citet[][$1.9\pm 0.1~\kpc$]{2017MNRAS.471.3057M}. However, if we fit the scale length to the spatial volumes with positive or negative $Y$-coordinates separately, we get very discrepant values of $h_{L,Y>0} = 2.7 \pm 0.2~\kpc$ and $h_{L,Y<0} = 1.9 \pm 0.1~\kpc$, showing how simplistic assumptions can produce biased results, which in this case was highly sensitive to the chosen spatial volume. This highlights the importance of focusing on accuracy and not only precision, given the enormous statistical power that is granted in the current \emph{Gaia} era. In order to not be mislead or misleading when it comes to measured properties of the Milky Way, it is crucial to not simply take such numbers at face value, but to carefully reflect on what those numbers mean in the context that they were produced and how they fit into the broader picture of ongoing and future analysis. In a similar vein, symmetry and equilibrium assumptions can bias measurements of for example stars' angle-action coordinates  (e.g. \citealt{2019ApJ...883..103B}) and the phase-space position of the Sun with respect to a Galactic rest-frame (e.g. \citealt{2016ARA&A..54..529B} discuss measurements and potential biases to the height of the Galactic mid-plane and the local standard of rest).

We plan to revisit and extend this analysis with future \emph{Gaia} data releases. The full third data release is planned for the first half year of 2022 and will contain 33 million radial velocity measurements\footnote{\url{https://www.cosmos.esa.int/web/gaia/dr3}} (compared to the current 7.6 million). Additionally, there are complementary distance information such as the photo-astrometric distance measurements produced with \texttt{StarHorse}, where \citet{2022A&A...658A..91A} claim a distance precision as good as 3~\%. With these improvements to the data, we will be able to apply this method with significantly greater depth, precision, and accuracy. This would give us a better understanding of the large-scale structure of the Galactic disk and the Milky Way's dynamics and history. This information will also propagate into other more global dynamical mass measurements (e.g. circular velocity curve analysis) and assist in constraining both the large-scale distribution of baryons as well as dark matter. In the longer term, we might be able to resolve variations in the Galactic potential also on smaller spatial scales, for example sourced by the Milky Way’s spiral arms or variations intrinsic to the nature of dark matter.

\section{Conclusion}\label{sec:conclusion}

This article is the third part of a longer series about a new method for weighing the Galactic disk, by using the time-varying dynamical structure of the phase-space spiral. In this work, we have applied our method to the \emph{Gaia} EDR3 proper motion sample in distant regions (1.4--3.4~kpc) of the Galactic disk. We can observe the spiral at this great depth without requiring radial velocity measurements, due to the fact that distant disk stars have a small Galactic latitude (to absolute value), such that their latitudinal proper motion has a close projection with vertical velocity.

This is the first analysis of its kind, in the sense that we are able to weigh the Galactic disk at large distances with a high spatial resolution. In our inference, we do not make strong assumptions about the spatial dependence of the Galactic disk surface density, for example in terms of axisymmetry or an exponential decay with Galactocentric radius. In our post-inference results, we do observe a decay of the disk surface density and fit a disk scale length of $2.2\pm 0.1~\kpc$. We also observe variations in the surface density on smaller spatial scales, of the order of 10--20~\%, which are possibly explained by systematic biases. Given these smaller scale variations, it's clear that a different spatial cut would give a significantly different result for the fitted scale length.

We plan to revisit the analysis made in this paper with future \emph{Gaia} data releases. This work is the first of its kind and although our analysis seems to suffer from some uncontrolled systematics of the order of roughly 10~\%, we have demonstrated that we can produce useful results for distant regions of the Galactic disk. We are confident that we will be able to improve and expand our analysis with \emph{Gaia}'s full third data release, also using supplementary photo-parallax information from the recently updated \texttt{StarHorse} catalogue \citep{2022A&A...658A..91A}, in order to go even further in distance and achieve greater accuracy.

\begin{acknowledgements}
AW acknowledges support from the Carlsberg Foundation via a Semper Ardens grant (CF15-0384).
CL acknowledges funding from the European Research Council (ERC) under the European Union's Horizon 2020 research and innovation programme (grant agreement
No. 852839).
GM acknowledges funding from the Agence Nationale de la Recherche (ANR project ANR-18-CE31-0006 and ANR-19-CE31-0017) and from the European Research Council (ERC) under the European Union's Horizon 2020 research and innovation programme (grant agreement No. 834148).
This work made use of an HPC facility funded by a grant from VILLUM FONDEN (projectnumber 16599).

This work has made use of data from the European Space Agency (ESA) mission \emph{Gaia} (\url{https://www.cosmos.esa.int/gaia}), processed by the \emph{Gaia} Data Processing and Analysis Consortium (DPAC,
\url{https://www.cosmos.esa.int/web/gaia/dpac/consortium}). Funding for the DPAC has been provided by national institutions, in particular the institutions participating in the \emph{Gaia} Multilateral Agreement.

This research utilised the following open-source Python packages: \textsc{Matplotlib} \citep{matplotlib}, \textsc{NumPy} \citep{numpy}, \textsc{SciPy} \citep{scipy}, \textsc{Pandas} \citep{pandas}, \textsc{TensorFlow} \citep{tensorflow2015-whitepaper}.
\end{acknowledgements}




\nocite{PaperOne}
\nocite{PaperTwo}
\nocite{PaperFour}
\bibliographystyle{aa} 
\bibliography{thisbib} 

\begin{appendix} 

\section{Coordinate transformations}
\label{app:coord_trans}

In this appendix, we describe how we transform the \emph{Gaia} observables to the solar rest-frame Cartesian coordinates $\boldsymbol{X}$ and $\boldsymbol{V}$, as defined in the beginning of Sect.~\ref{sec:definitions}. The \emph{Gaia} observables include the parallax $\varpi$, the Galactic longitude and latitude ($l$ and $b$), their corresponding proper motions ($\mu_l$ and $\mu_b$), and the radial velocity ($v_\text{RV}$).

The spatial position is given by
\begin{equation}
    \boldsymbol{X} =
    \begin{bmatrix}
    \cos l \times \cos b \\
    \sin l \times \cos b \\
    \sin b
    \end{bmatrix}
    \times \frac{\mas\; \kpc}{\varpi}.
\end{equation}
The height with respect to the Galactic mid-plane is then found via a translation according to Eq.~\eqref{eq:Z_to_z}.

The velocities in the solar rest frame are given by
\begin{equation}
    \boldsymbol{V} = \boldsymbol{R}(l,b) \times
    \begin{bmatrix}
    k_\mu \times \mu_l / \varpi \\
    k_\mu \times \mu_b / \varpi \\
    v_\text{RV}
    \end{bmatrix},
\end{equation}
where $k_\mu = 4.74057 ~ \yr \times \kmsec$ and
\begin{equation}\label{eq:rotation_matrix}
	\boldsymbol{R}(l,b) =
    \begin{bmatrix}
    -\sin l &  -\cos l \times \sin b & \cos l \times \cos b \\
    \cos l & -\sin l \times \sin b & \sin l \times \cos b  \\
    0 & \cos b & \sin b
    \end{bmatrix}
\end{equation}
is a rotational matrix. The vertical velocity in the rest frame of the Galactic disk is then found via a translation according to Eq.~\eqref{eq:W_to_w}.

There is further important information on how the data is processed, such as accounting for the \emph{Gaia} EDR3 zero-point offset and assigning missing radial velocities. This is described in detail in Sect.~\ref{sec:data}.

\section{Supplementary plots}
\label{app:suppl_plots}

\begin{figure}
	\includegraphics[width=1.\columnwidth]{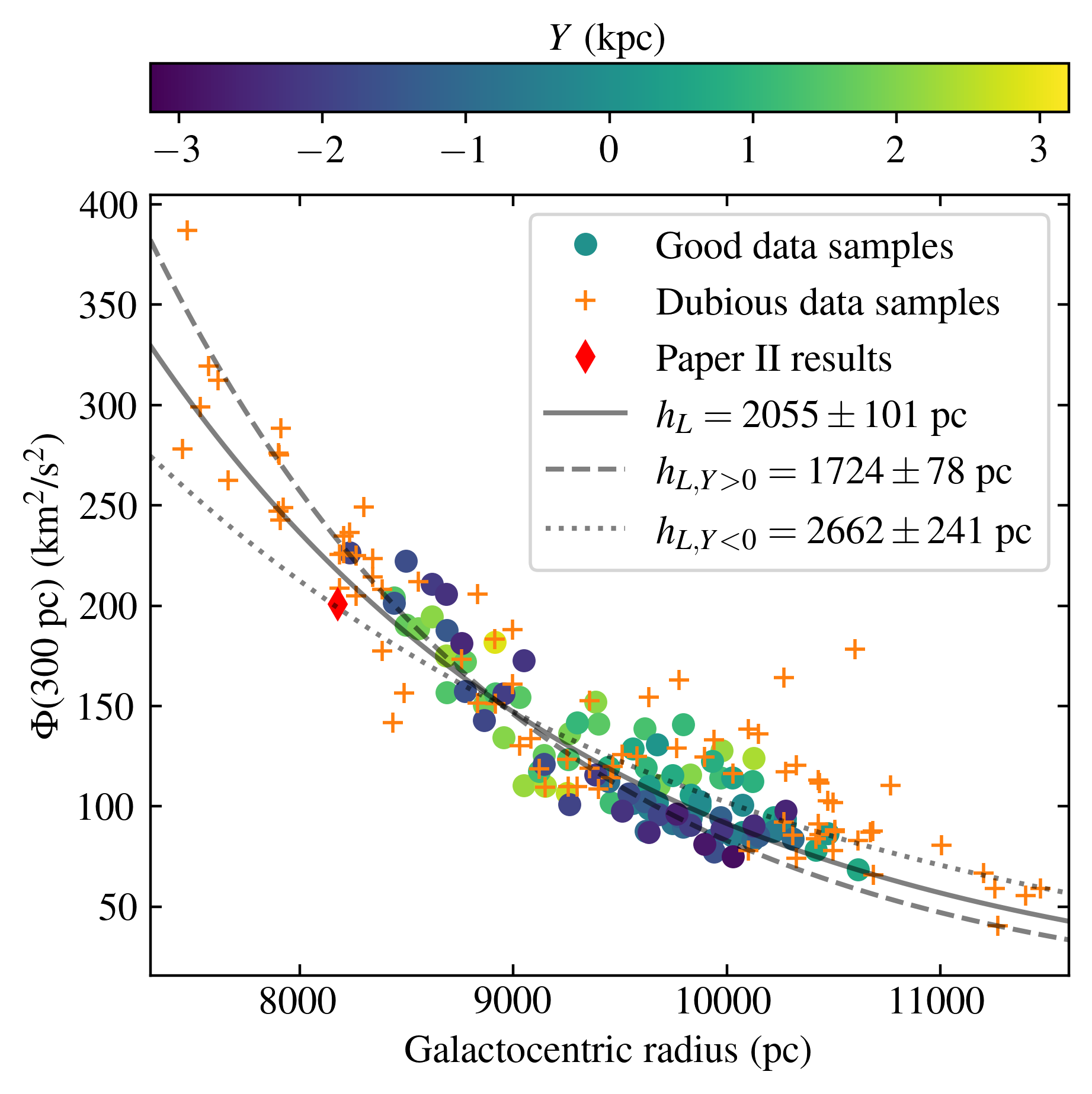}
    \caption{Same as Fig.~\ref{fig:phi_of_R_400pc}, but for $|z|=300~\pc$.}
    \label{fig:phi_of_R_300pc}
\end{figure}

\begin{figure}
	\includegraphics[width=1.\columnwidth]{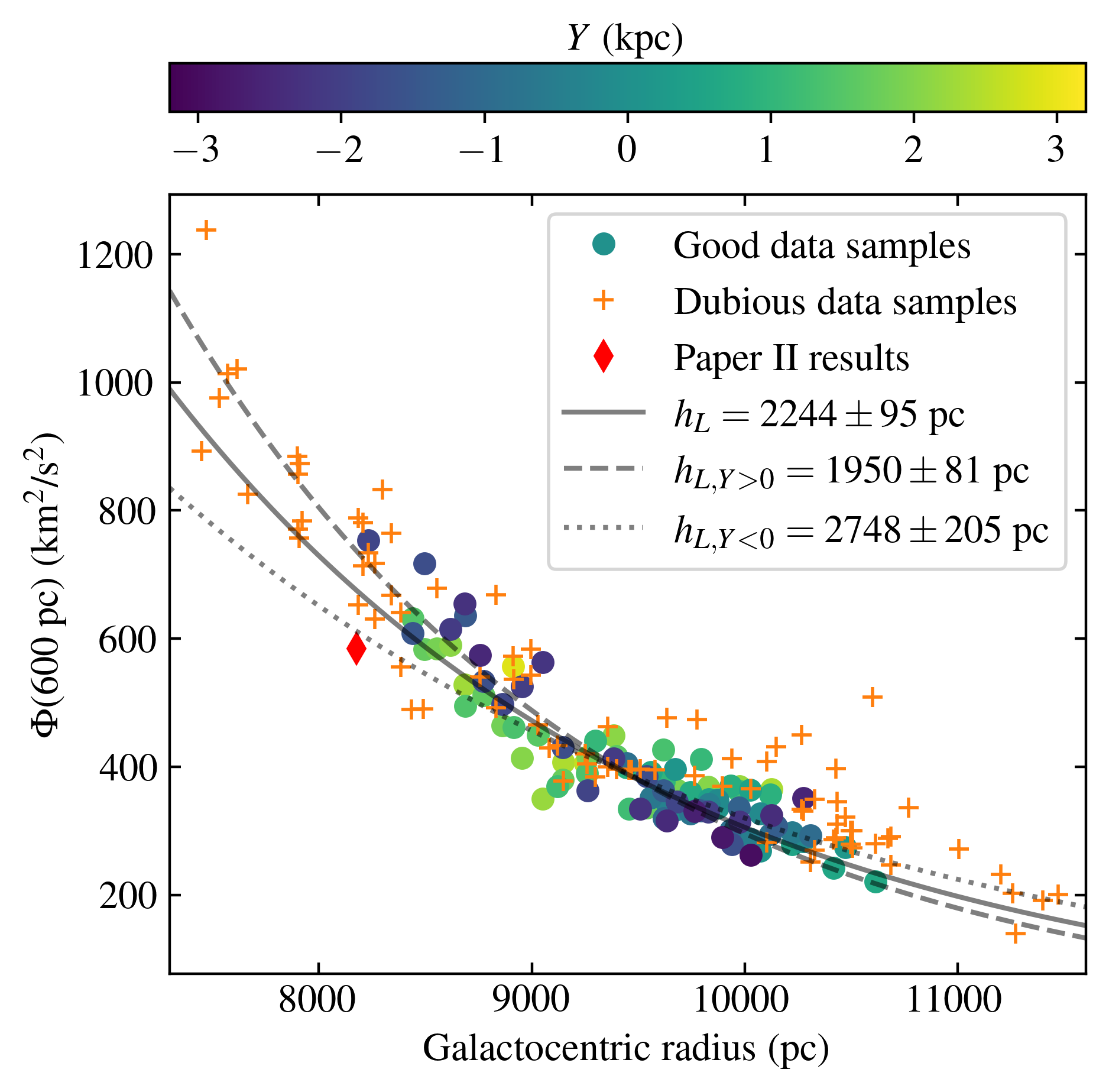}
    \caption{Same as Fig.~\ref{fig:phi_of_R_400pc}, but for $|z|=600~\pc$.}
    \label{fig:phi_of_R_600pc}
\end{figure}

\begin{figure}
	\includegraphics[width=1.\columnwidth]{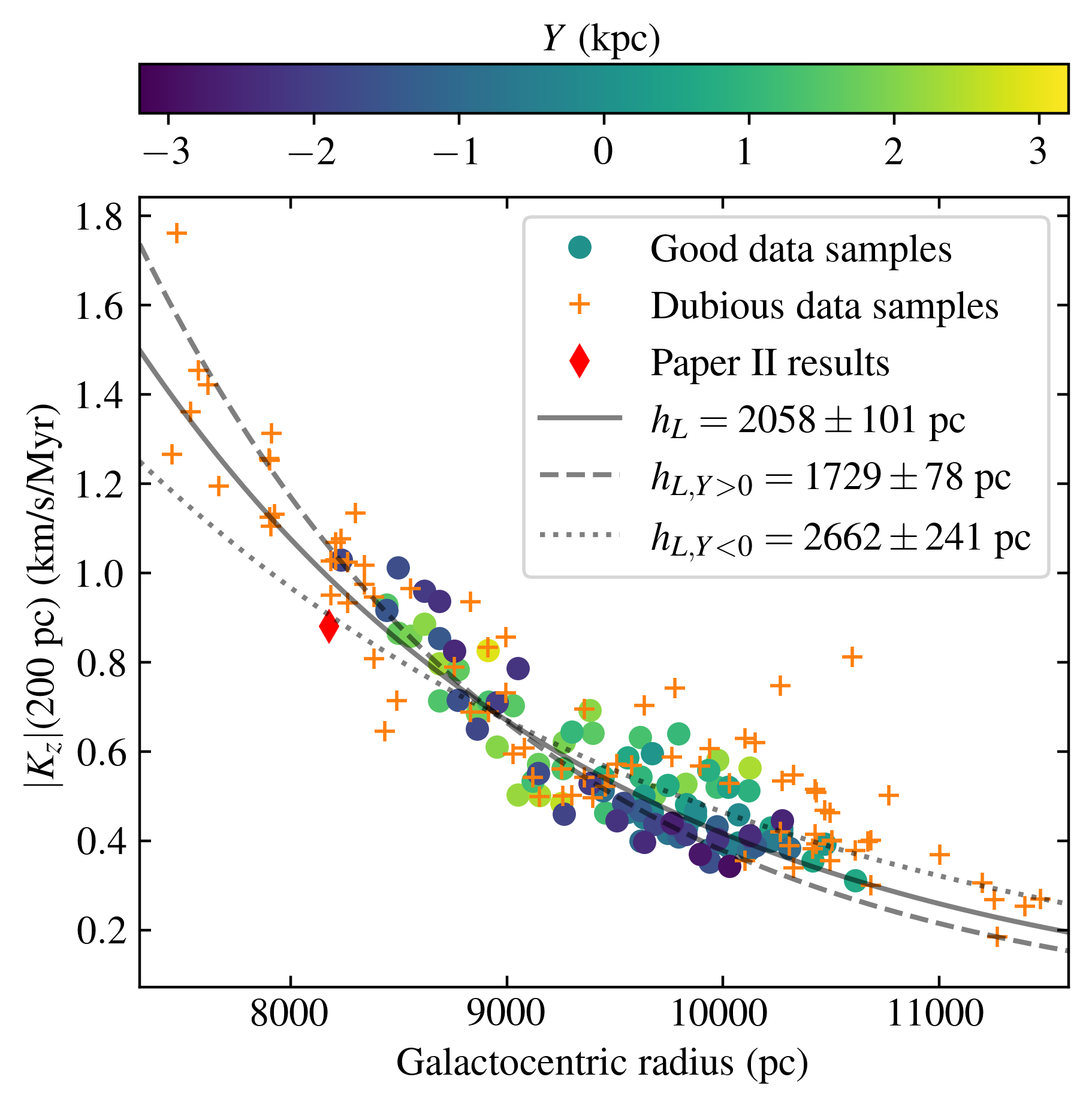}
    \caption{Same as Fig.~\ref{fig:phi_of_R_400pc}, but for the vertical acceleration at $|z|=200~\pc$.}
    \label{fig:Kz_of_R_200pc}
\end{figure}

\begin{figure}
	\includegraphics[width=1.\columnwidth]{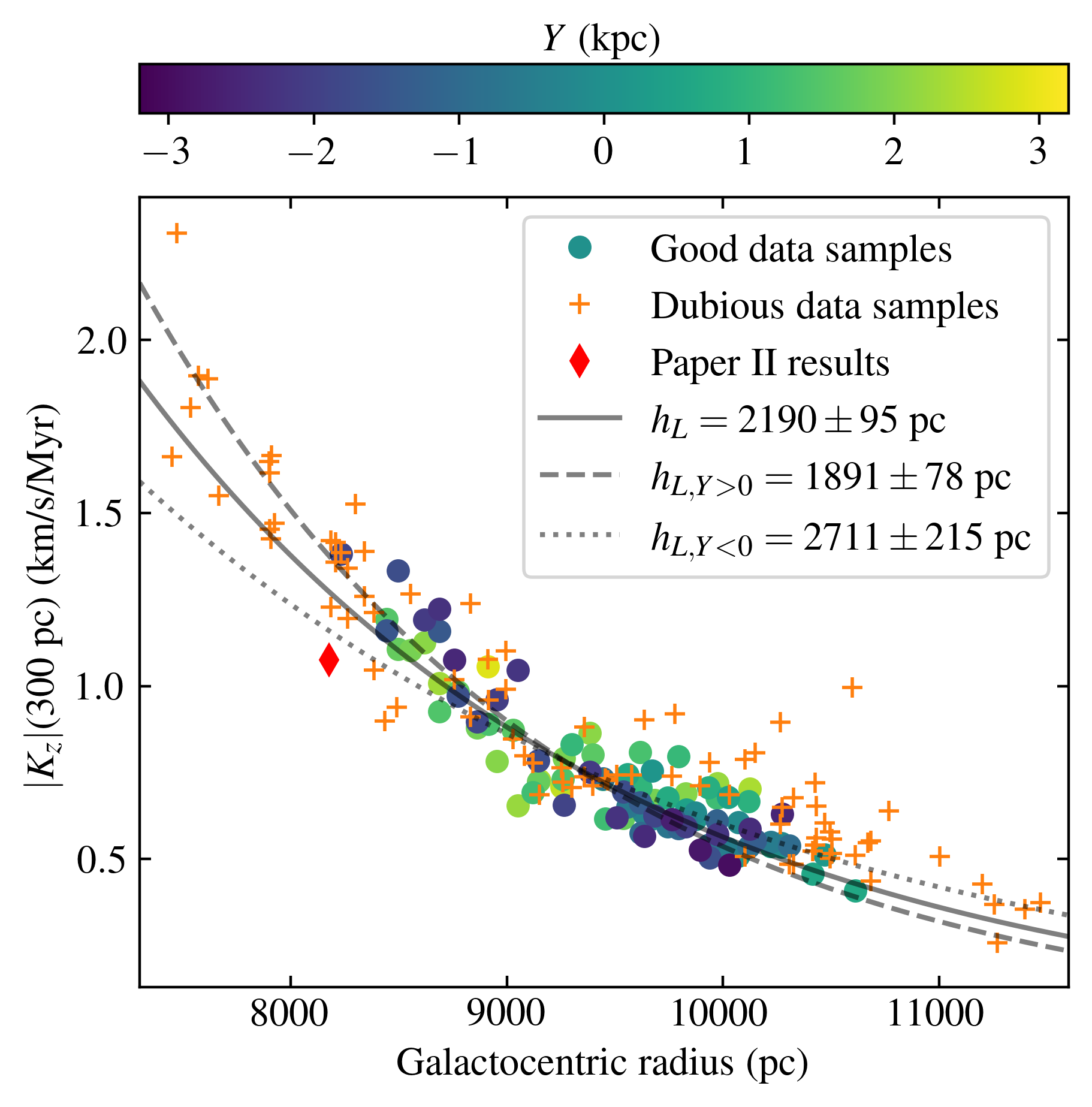}
    \caption{Same as Fig.~\ref{fig:phi_of_R_400pc}, but for the vertical acceleration at $|z|=300~\pc$.}
    \label{fig:Kz_of_R_300pc}
\end{figure}

\begin{figure}
	\includegraphics[width=1.\columnwidth]{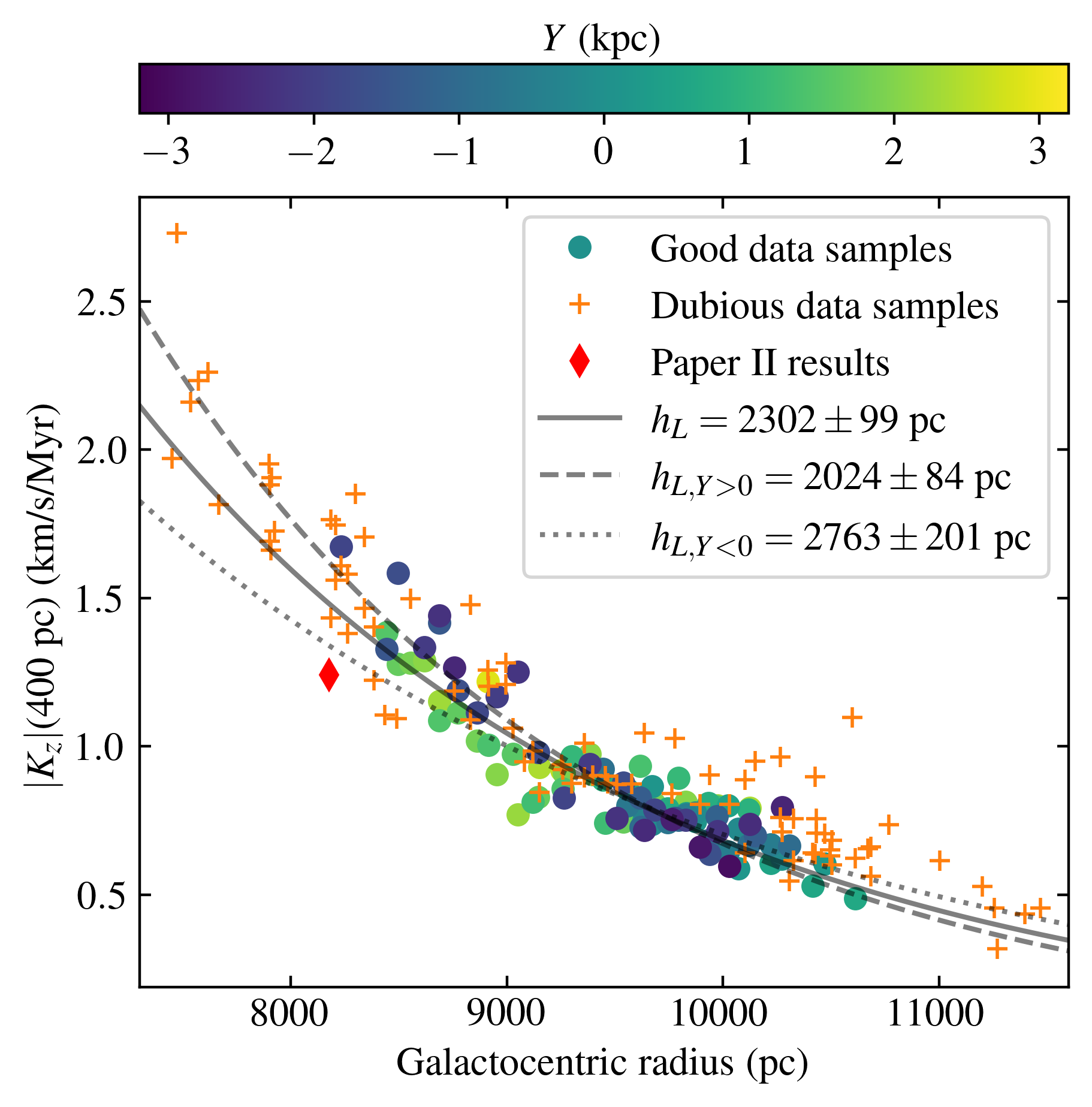}
    \caption{Same as Fig.~\ref{fig:phi_of_R_400pc}, but for the vertical acceleration at $|z|=400~\pc$.}
    \label{fig:Kz_of_R_400pc}
\end{figure}

In this appendix, we show some additional plots of our results. In Figs.~\ref{fig:phi_of_R_300pc}--\ref{fig:Kz_of_R_400pc}, we show how the inferred vertical gravitational potential values $\Phi(300~\pc)$ and $\Phi(600~\pc)$, as well as the inferred vertical acceleration values $K_z(200~\pc)$, $K_z(300~\pc)$, and $K_z(400~\pc)$, depend on Galactocentric radius. Their azimuthal dependence is also visible via the marker colour coding. In all these figures, we see very similar disk scale lengths and a similar general behaviour when splitting the data samples into two separate groups along the solar azimuth.

\begin{figure}
	\includegraphics[width=1.\columnwidth]{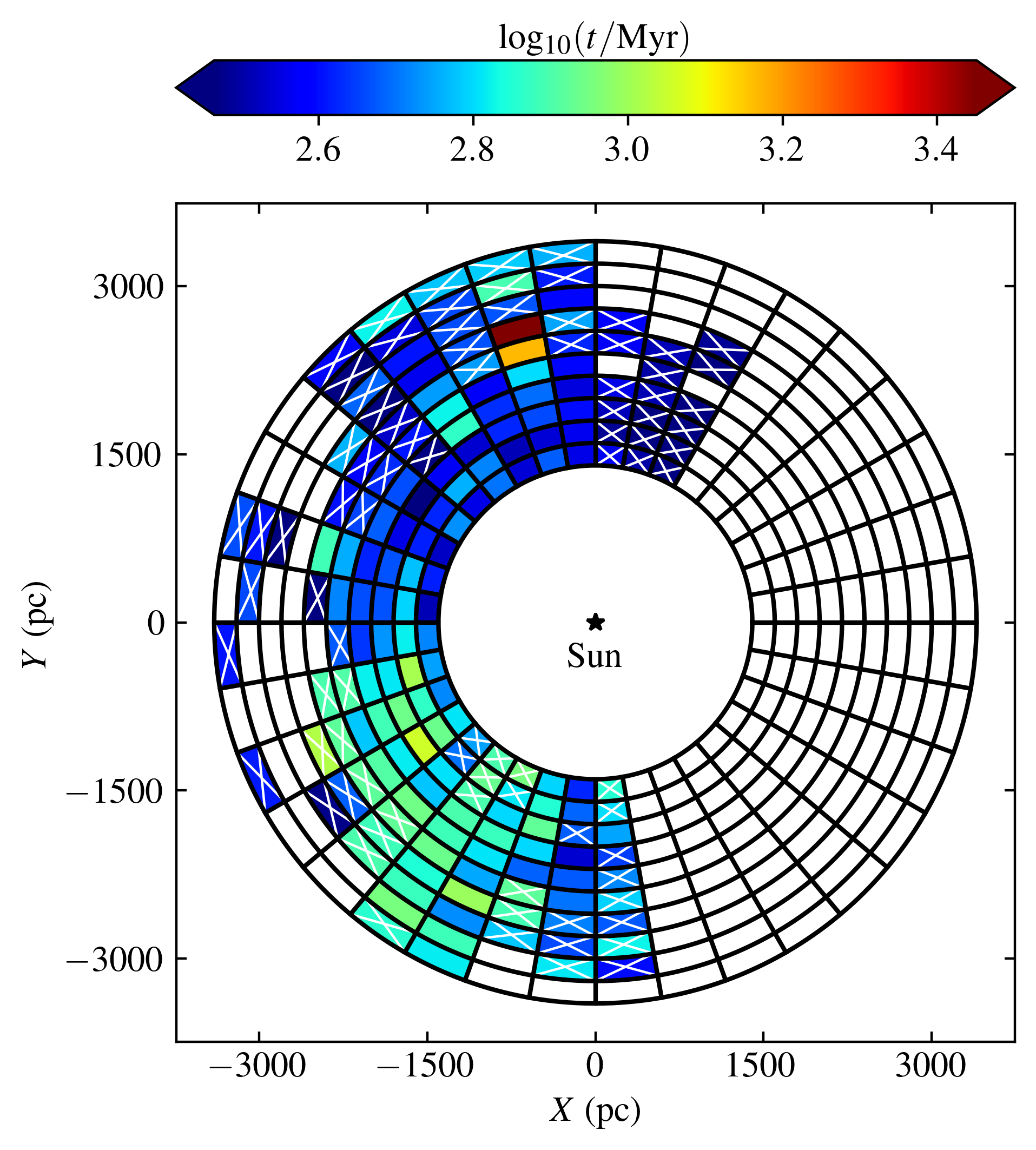}
    \caption{Inferred time since the perturbation was produced (model parameter $t$) for the respective data samples. Disqualified data samples are left blank and dubious data samples are marked with a white cross.}
    \label{fig:t_pert}
\end{figure}

In Fig.~\ref{fig:t_pert} we show the inferred time since the perturbation (the model parameter $t$). As we saw in our tests on simulations in \citetalias{PaperI}, this parameter is not very robustly inferred, due to its strong degeneracy with respect to the precise shape of the gravitational potential. We do in fact see a manifestation of this degeneracy, in that the inferred time has a strong correlation with the broken axisymmetry of the inferred gravitational potential at low heights (seen most clearly in $\Phi(300~\pc)$). This is further indication that the observed broken axisymmetry is likely due to a systematic bias. We also see two very strong outliers in the direction of $l\in [100,110]~\deg$. The inferred gravitational potentials of these two data samples are close to harmonic, which pushes the inferred time to high values. Due to these two outliers, the distribution of inferred times is best captured by its 16th, 50th, and 84th percentiles; for our good data samples, they are 350, 622, and 767~Myr, respectively. This is in decent agreement with the results of for example \cite{2018Natur.561..360A} and \cite{2019MNRAS.484.1050D}.

\section{Examples of dubious data samples}\label{app:examples}

In this appendix, we show a few examples of dubious data samples, in terms of their data histograms and inferred phase-space densities. This is done in order to illustrate our rationale for disqualifying and marking data samples as dubious.

\begin{figure*}
	\includegraphics[width=1.\textwidth]{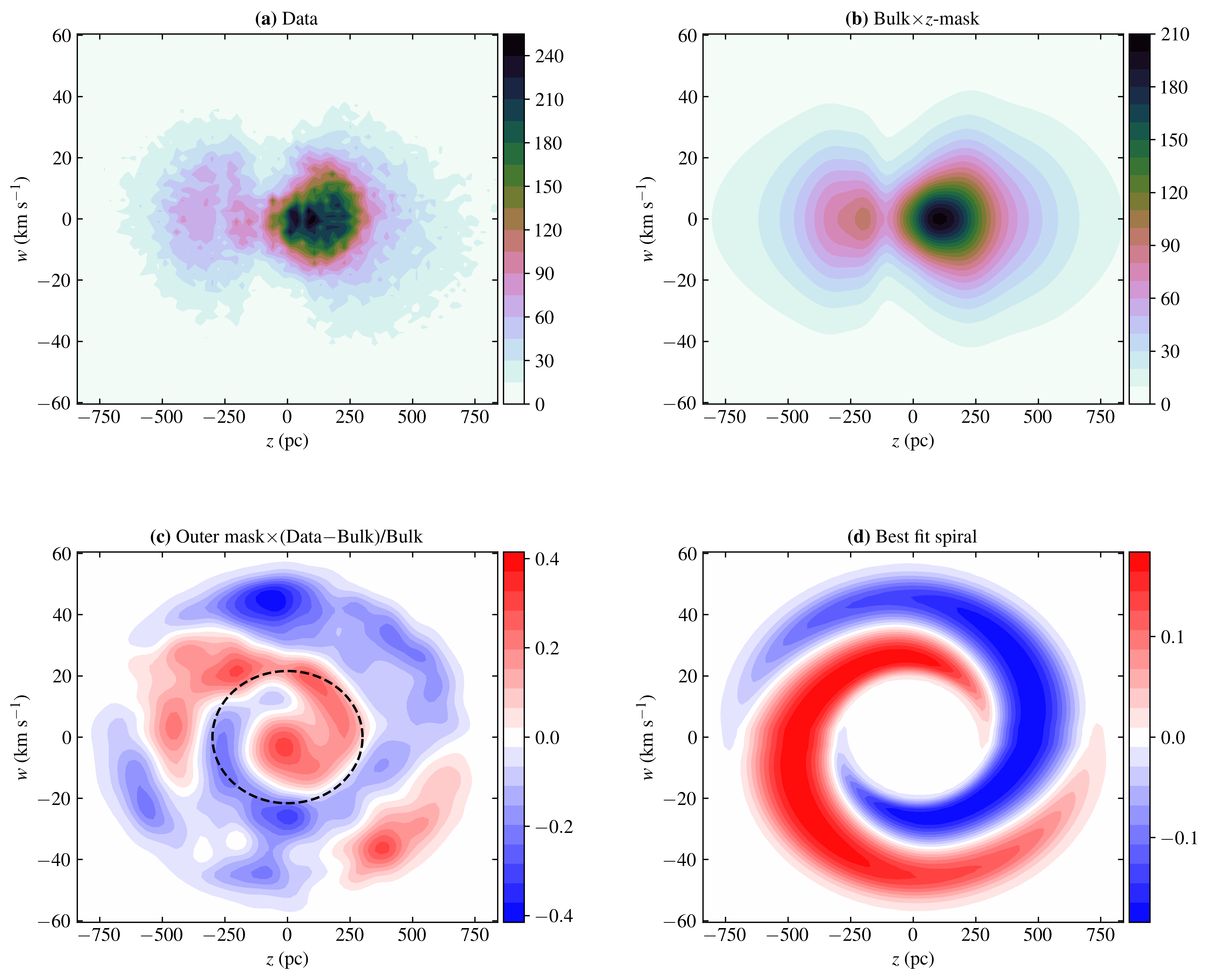}
    \caption{Same as Fig.~\ref{fig:spiral_illustrative_example}, but for the data sample with $l \in [60,70]~\deg$ and $\sqrt{X^2+Y^2} \in [2000,2200]~\pc$. This data sample was marked as dubious.}
    \label{fig:spiral_skewed_example}
\end{figure*}

In Fig.~\ref{fig:spiral_skewed_example}, we show the data histogram and inferred phase-space density of the data sample with $l \in [60,70]~\deg$ and $\sqrt{X^2+Y^2} \in [2000,2200]~\pc$. This data sample is representative of the larger spatial region of $l \in [60,90]~\deg$, which was all disqualified or marked as dubious. The data samples in this region suffer from fairly strong extinction but also a skewed spiral, as seen in panel {\bf (c)} of this figure. The skewed shape of the spiral is likely produced by some systematic bias, plausible related to the assignment of radial velocities, as that could indeed skew the spiral in this manner (see Fig.~\ref{fig:spiral_RV_bias}). Even though this region ($l \in [60,90]~\deg$) had no good data samples, its dubious data samples still produced reasonable results that agreed well with our inferred disk scale length (see especially Figs.~\ref{fig:phi_of_R_400pc} and \ref{fig:phi_of_R_500pc}).

\begin{figure*}
	\includegraphics[width=1.\textwidth]{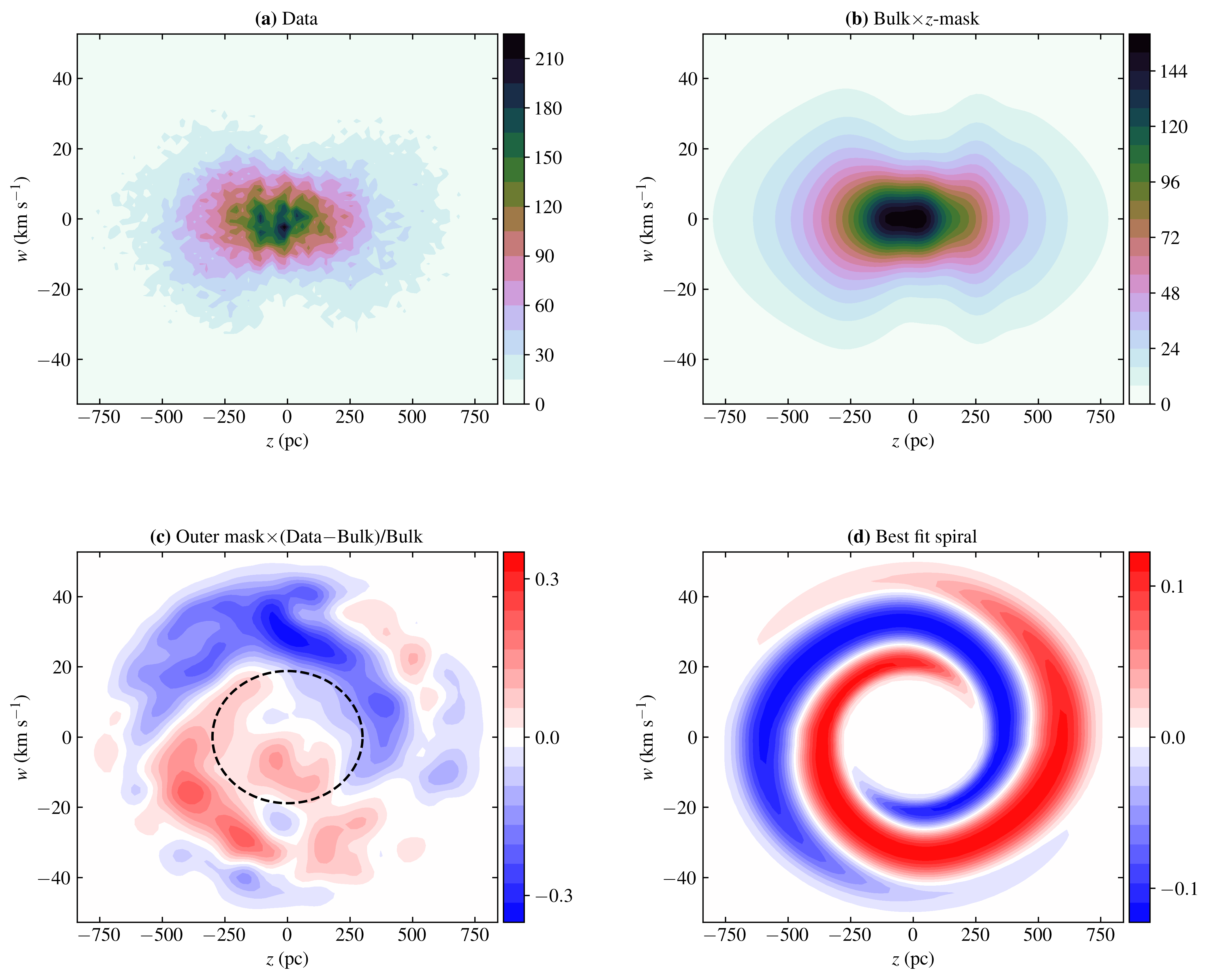}
    \caption{Same as Fig.~\ref{fig:spiral_illustrative_example}, but for the data sample with $l \in [270,280]~\deg$ and $\sqrt{X^2+Y^2} \in [2200,2400]~\pc$. This data sample was marked as dubious.}
    \label{fig:spiral_asymm_example}
\end{figure*}

In Fig.~\ref{fig:spiral_asymm_example}, we show the data histogram and inferred phase-space density of the data sample with $l \in [270,280]~\deg$ and $\sqrt{X^2+Y^2} \in [2200,2400]~\pc$, which is representative of the region with the same Galactic longitude. Looking at panel {\bf (c)}, the phase-space spiral is clearly present in the data as a perturbation with respect to the fitted bulk density. However, the velocity distribution does seem rather asymmetric and skewed (in addition to the spiral perturbation), with a very red (blue) region at the bottom (top) of the panel. If this is due to an actual skewed velocity distribution or some systematic bias is difficult to say, but either way it lead us to mark the data samples in this region as dubious. Again, most data samples in this region agree fairly well with the general trends of our results.

\begin{figure*}
	\includegraphics[width=1.\textwidth]{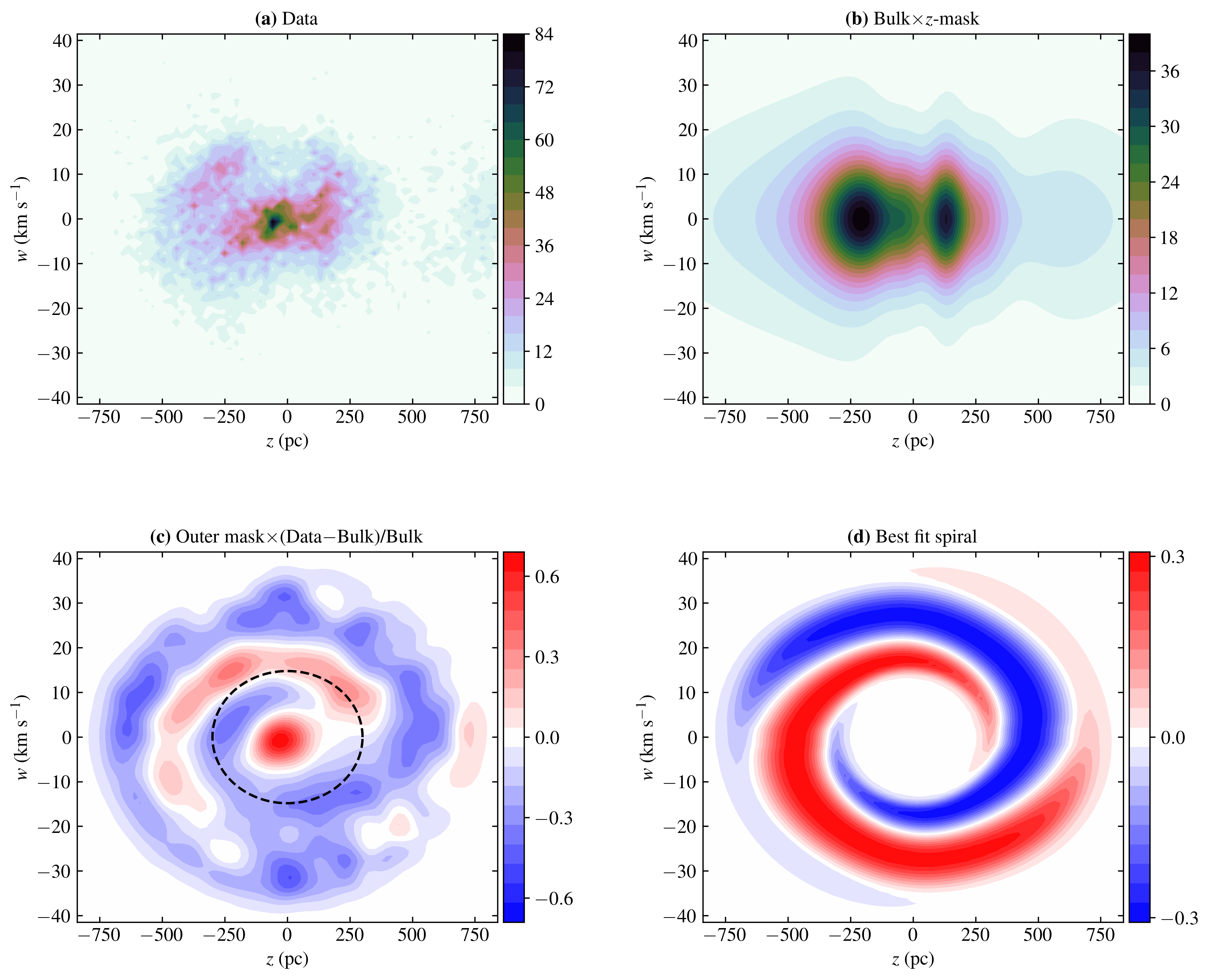}
    \caption{Same as Fig.~\ref{fig:spiral_illustrative_example}, but for the data sample with $l \in [130,140]~\deg$ and $\sqrt{X^2+Y^2} \in [2800,3000]~\pc$. This data sample was marked as dubious.}
    \label{fig:spiral_extinction_example}
\end{figure*}

In Fig.~\ref{fig:spiral_extinction_example}, we show the data histogram and inferred phase-space density of the data sample with $l \in [130,140]~\deg$ and $\sqrt{X^2+Y^2} \in [2800,3000]~\pc$. This is an example of a data sample which overall seemed well behaved but suffered from fairly strong selection effects. The inferred spiral seen in panel {\bf (d)} seems like a good fit, but the spiral in panel {\bf (c)} is not seen as a clearly continuous structure. This lead us to mark this data sample as dubious.

\end{appendix}

\end{document}